\documentclass[onecolumn,useAMS,usenatbib]{mn2e}

\usepackage{graphicx,amsmath,color}

\usepackage{natbib}

\usepackage{bm}

\title[GII self-calibration]
{Self-calibrating the gravitational shear-intrinsic ellipticity-intrinsic ellipticity (GII) cross-correlation}

\author[Troxel \& Ishak]
 {M. A. Troxel\thanks{Electronic address:
    {\tt troxel@utdallas.edu}} and
  M. Ishak\thanks{Electronic address:
    {\tt mishak@utdallas.edu}}$^1$ 
\\$^1$Department of Physics, The University of Texas at Dallas, Richardson, TX 75083, USA
}

\begin{document}

\date{\today}

\pagerange{\pageref{firstpage}--\pageref{lastpage}} \pubyear{0000}

\maketitle

\label{firstpage}

\begin{abstract}
We extend the 3-point intrinsic alignment self-calibration technique to the gravitational shear-intrinsic ellipticity-intrinsic ellipticity (GII) bispectrum. The proposed technique will allow the measurement and removal of the GII intrinsic alignment contamination from the cross-correlation weak lensing signal. While significantly decreased from using cross-correlations instead of auto-correlation in a single photo-z bin, the GII contamination persists in adjacent photo-z bins and must be accounted for and removed from the lensing signal. We relate the GII and galaxy density-intrinsic ellipticity-intrinsic ellipticity (gII) bispectra through use of the galaxy bias, and develop the estimator necessary to isolate the gII bispectrum from observations. We find that the GII self-calibration technique performs at a level comparable to that of the gravitational shear-gravitational shear-intrinsic ellipticity correlation (GGI) self-calibration technique, with measurement error introduced through the gII estimator generally negligible when compared to minimum survey error. The accuracy of the relationship between the GII and gII bispectra typically allows the GII self-calibration to reduce the GII contamination by a factor of 10 or more for all adjacent photo-z bin combinations at $\ell>300$. For larger scales, we find that the GII contamination can be reduced by a factor of 3-5 or more. The GII self-calibration technique is complementary to the existing GGI self-calibration technique, which together will allow the total intrinsic alignment cross-correlation signal in 3-point weak lensing to be measured and removed.
\end{abstract}

\begin{keywords}
gravitational lensing -- cosmology 
\end{keywords}

\section{Introduction}\label{GIIintro}

Weak gravitational lensing due to large scale structure (cosmic shear) has become a promising source of cosmological information. A new generation of ground- and space-based surveys suited for precision weak lensing measurements have been developed with the importance of this new probe in mind. These ongoing, future, and proposed surveys (e.g. CFHTLS\footnote{{\slshape http://www.cfht.hawaii.edu/Science/CFHLS/}}, DES\footnote{{\slshape http://www.darkenergysurvey.org/}}, EUCLID\footnote{{\slshape http://sci.esa.int/euclid/}}, HSC\footnote{{\slshape http://www.naoj.org/Projects/HSC/}}, HST\footnote{{\slshape http://www.stsci.edu/hst/}}, JWST\footnote{{\slshape http://www.jwst.nasa.gov/}}, LSST\footnote{{\slshape http://www.lsst.org/lsst/}}, Pan-STARRS\footnote{{\slshape http://pan-starrs.ifa.hawaii.edu/}}, and WFIRST\footnote{{\slshape http://wfirst.gsfc.nasa.gov/}}) promise to provide greatly improved measurements of cosmic shear using the shapes of up to billions of galaxies. There has been much work done to explore the potential of these cosmic shear measurements, which we review in \cite{troxel}, for both the 2- and 3-point cosmic shear correlations.

Beyond the constraints obtained on cosmological parameters from the 2-point cosmic shear correlation and the corresponding shear power spectrum, the 3-point cosmic shear correlation and shear bispectrum are able to break degeneracies between the cosmological parameters that the power spectrum alone does not \citep{3a,3b}. The results of \cite{tj}, for example, showed that the constraints on the dark energy parameters and the matter fluctuation amplitude should be able to be improved by a further factor of 2-3 using the bispectrum measured in a deep lensing survey. Most recently, parameter constraints were derived by \cite{5} using weak lensing data from the HST COSMOS survey, measuring the third order moment of the aperture mass measure. Their independent results were consistent with WMAP7 best-fit cosmology and provided an improved constraint when combined with the 2-point correlation. The bispectrum also allows us to explore information about non-Gaussianity in the universe that is inaccessible at the 2-point level, providing constraints on the degree of non-Gaussianity in addition to the information on other cosmological parameters (see for example \cite{matarrese,verde,tj,28,huterer,munshi} and references therein.) 

However, several systematic effects limit the precision of cosmic shear measurements, which must be accounted for in order to make full use of the potential of future weak lensing surveys (see for a summary \cite{troxel} and references therein), and one of the serious systematic effects of weak lensing is this correlated intrinsic alignment of galaxy ellipticities, which act as a nuisance factor (see for example \cite{6d,6c,8c,7,8g,6j,8e,8f,18b,hirata04,17,13,12,bk,10,semb,14a,14b,9b,15,19,8a,8h,troxel2} and references therein). The dark energy equation of state can be biased by as much as $50\%$, for example, if intrinsic alignment is ignored \citep{bk,9b}. \cite{10} found that the matter power spectrum amplitude can be affected by up to $30\%$ due to the intrinsic alignment, demonstrating the importance of developing methods to isolate and remove the intrinsic alignment from the cosmic shear signal.

\begin{figure}
\center
\includegraphics[scale=0.6]{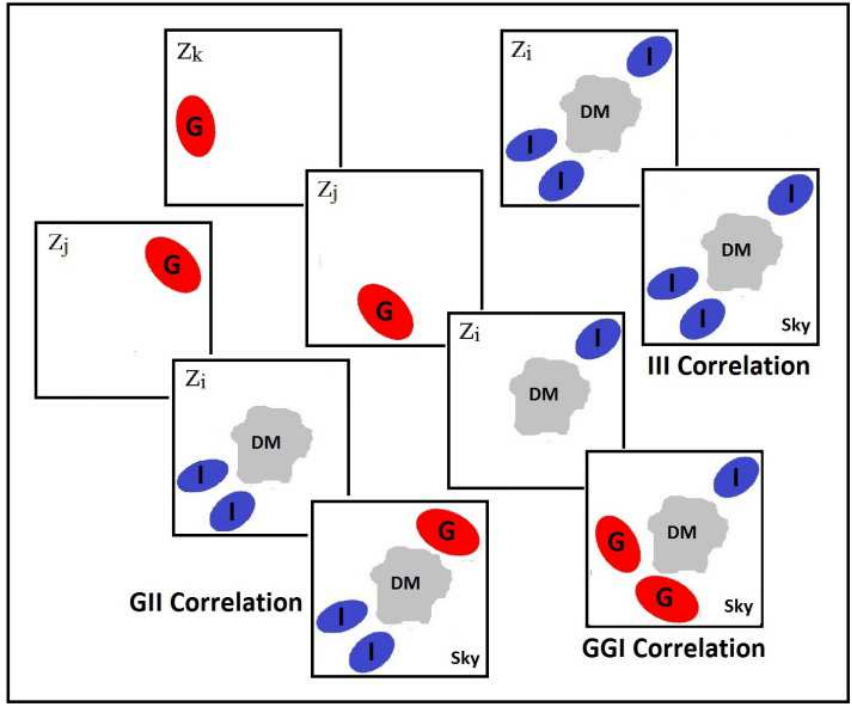}%
\caption{The 3-point galaxy intrinsic alignment correlations. Blue galaxies are intrinsically aligned (I), while red galaxies are lensed (G). The view of the system on the sky is represented in the lower right panels. Each preceding panel demonstrates the galaxy configuration at some distinct redshift such that $z_i<z_j<z_k$.}\label{GIIfig:3ptia}
\end{figure}

There are two 2-point intrinsic alignment correlations. The first is a correlation between the intrinsic ellipticity of two galaxies, known as the intrinsic ellipticity-intrinsic ellipticity (II) correlation, between two intrinsically aligned galaxies. The second, known as the gravitational shear-intrinsic ellipticity (GI) correlation, was identified by \cite{hirata04} and is due to a matter structure both causing the alignment of a nearby galaxy and contributing to the lensing signal of a background galaxy. The large-scale GI correlation was first reported by \cite{12} in the SDSS, with \cite{10} finding an even stronger GI correlation for Luminous Red Galaxies (LRGs). The authors showed that this contamination can affect the lensing measurement and cosmology by up to $10\%$, while affecting the matter fluctuation amplitude by up to $30\%$. \cite{13} confirmed this through numerical simulations, where a level of contamination of $10\%$ was found. The GI correlation was also measured in the SDSS dataset by \cite{14a} and \cite{14b}, and most recently, \cite{15} measured strong 2-point intrinsic alignment correlations in various SDSS and MegaZ-LRG samples.

Similarly, when we consider the 3-point correlation, the gravitational shear-gravitational shear-gravitational shear (GGG) bispectrum suffers from contamination by the 3-point intrinsic alignment correlations. The first correlation is between the intrinsic ellipticities of three spatially close galaxies which are intrinsically aligned by a nearby matter structure, known as the intrinsic ellipticity-intrinsic ellipticity-intrinsic ellipticity (III) correlation. The second is the gravitational shear-intrinsic ellipticity-intrinsic ellipticity (GII) correlation, where a matter structure which contributes to the lensing of a galaxy in the background also intrinsically aligns two spatially nearby galaxies. Finally, there is the gravitational shear-gravitational shear-intrinsic ellipticity (GGI) correlation, where a matter structure which intrinsically aligns a galaxy in the foreground also lenses two galaxies in the background. The sign of the GGI and GII correlations can depend both on triangle shape and scale. The 3-point intrinsic alignment correlations are represented diagrammatically in Fig. \ref{GIIfig:3ptia}. It was shown by \cite{semb} that compared to the lensing spectrum, the lensing bispectrum is typically more strongly contaminated by galaxy intrinsic alignment, up to $15-20\%$ compared to the GGG lensing signal. 

\cite{troxel3} demonstrated that the III and GII intrinsic alignment correlations can be greatly reduced with photo-z's by using cross-spectra of galaxies in two different redshift bins so that the galaxies are separated by large enough distances to assure that the tidal effect is weak. However, as shown in Fig. 4 of \cite{troxel3}, while the III bispectrum is supressed almost totally for adjacent, large redshift bins, the GII bispectrum is not. Thus we propose an extension to the GGI self-calibration technique \citep{troxel} for the GII bispectrum below. There has been significant work in recent years toward measuring and removing the effect of galaxy intrinsic alignment on cosmological measurements, and we review these briefly. The 2-point intrinsic alignment correlations and methods for their removal have been well studied in recent literature (see for example \cite{hirata04,18b,17,bk,9b,19,20a,20b,20c,22,23}). Using a geometrical approach, \cite{21} generalized the nulling technique to the 3-point GGI correlation by exploiting its redshift dependence, but it was found that the technique throws out some of the valuable lensing signal. Finally, we have recently generalized the approaches of \cite{22} and \cite{23} to develop a set of 3-point self-calibration techniques. We proposed using additional galaxy density (cross-)correlations which are already present in weak lensing survey measurements to self-calibrate the GGI cross-correlation in \cite{troxel}, while the unique redshift dependencies of the intrinsic alignment bispectra allowed us to self-calibrate the intrinsic alignment auto-correlations in \cite{troxel3}.

We propose here an extension to the GGI self-calibration technique of \cite{troxel}, developing a technique to self-calibrate the GII cross-correlation bispectrum. While the impact of the GII cross-correlation on cosmological information is significantly less than the impact of the GGI cross-correlation, especially for widely separated redshift bin combinations, it still has a contribution for adjacent redshift bin combinations, which must be measured and removed if one hopes to achieve precision, systematic free bispectrum measurements. To do this, we develop a relationship between the GII and galaxy density-intrinsic ellipticity-intrinsic ellipticity (gII) bispectrum. We also devise an estimator to measure the gII bispectrum from the observed galaxy density-ellipticity-ellipticity bispectrum. We then discuss the performance of such a self-calibration technique, describing both measurement and systematic errors due to the self-calibration.

We organize the paper as follows. In Sec. \ref{GIIback}, we briefly discuss the necessary survey parameters and lensing calculations. In Sec. \ref{GIIsc}, we develop the 3-point GII self-calibration. We first develop a relation between the GII and gII bispectra, which allows for the estimation and removal of the GII intrinsic alignment cross-correlation from the cosmic shear signal. We then establish an estimator to extract the galaxy density-density-intrinsic ellipticity correlation (ggI) from the observed galaxy ellipticity-density-density measurement for a photo-z galaxy sample. Section \ref{GIIerror} describes the residual sources of error to the GII self-calibration technique, and we present the necessary relations to quantify these errors. Finally, we summarise the effectiveness and impact of the GII self-calibration in Sec. \ref{GIIconc}. We expand in the Appendix upon the detailed calculation of the coefficients in the error calculation found in Sec. \ref{GIIciggerror} and provide a list of expected values.

\section{Background}\label{GIIback}

The self-calibration technique proposed by \cite{22} for the 2-point intrinsic alignment correlations and generalized by \cite{troxel} for the 3-point correlations makes use of the information already found in a lensing survey, including galaxy shape, angular position and photometric redshift, in order to calculate and remove the dominant intrinsic alignment contaminations. In evaluating the performance of the self-calibration technique, we will consider as an example survey parameters to match an LSST-like weak lensing survey \citep{lsst}, but of course the calculations are applicable to all current and planned weak lensing surveys (e.g. CFHTLS, DES, EUCLID, HSC, HST, JWST, LSST, Pan-STARRS, and WFIRST). We divide galaxies into photo-z bins according to photo-z $z^P$, where the \emph{i}-th photo-z bin has a range $\bar{z}_i-\Delta z_i/2\le z^P\le \bar{z}_i+\Delta z_i/2$ for mean photo-z $\bar{z}_i$. In this notation, $i<j$ implies that $\bar{z}_i<\bar{z}_j$. The galaxy redshift distribution over the \emph{i}-th redshift bin is $n^P_i(z^P)$ and $n_i(z)$ as a function of photo-z and true redshift, respectively. These are then related by a photo-z probability distribution function (PDF) $p(z|z^P)$. 

Our calculations assume that the survey will acheive a half sky coverage ($f_{sky}=0.5$) and a galaxy surface density of 40 arcminute$^{-2}$ with redshift density distribution of
\begin{equation}
n(z)=\frac{1}{2z_0}\left(\frac{z}{z_0}\right)^2\exp(-z/z_0),
\end{equation}
with $z_0=0.5$. The ellipticity shape noise is described by $\gamma_{rms}=0.18+0.042z$ \citep{22} and the photo-z error by a Gaussian PDF of the form
\begin{equation}
p(z|z^P)=\frac{1}{\sqrt{2\pi\sigma_z^2}}\exp\left(\frac{-(z-z^P)^2}{2\sigma_z^2}\right),
\end{equation}
with $\sigma_z=0.05(1+z)$. We construct photo-z redshift bins with width $\Delta z=0.2$, centred at $\bar{z}_i=0.2(i+1)$ ($i=1,\cdots,9$). Redshifts below $z^P=0.3$ are excluded, not because of poor performance in the GII self-calibration technique, but rather due to the weaker lensing signal at low redshifts, which artificially inflates the fractional errors we evaluate in Sec. \ref{GIIerror}. These errors are measured with respect to the lensing signal, and thus lower photo-z bins are not useful in evaluating the true performance of the GII self-calibration.

The intrinsic alignment self-calibration techniques (GII and GGI) both rely upon two basic observables measured in a weak lensing survey: galaxy surface density and galaxy shape. The galaxy surface density, $\delta^{\Sigma}$, is a function of the 3D galaxy distribution $\delta_g$ in a given photo-z bin. The galaxy shape is expressed in terms of ellipticity, which measures the cosmic shear $\gamma$. However, the intrinsic ellipticities of galaxies contaminate the cosmic shear. This intrinsic ellipticity has a random component, which is simple to correct for and which we include as part of the shot noise in the error estimations of Sec. \ref{GIIerror}. A second component to this intrinsic ellipticity is due to the correlated intrinsic alignment of galaxies and was introduced in Sec. \ref{GIIintro}. The measured shear can be labelled as $\gamma^s=\gamma+\gamma^I$, where $\gamma^I$ denotes the correlated intrinsic ellipticity due to intrinsic galaxy alignment. We are concerned only with the weak limit of gravitational lensing, so we can work with the lensing convergence $\kappa$ instead. Thus from the measured $\gamma^s$, we can obtain the convergence $\kappa^s=\kappa+\kappa^I$.

We assume a standard, flat $\Lambda$CDM universe in our calculations. The convergence $\kappa$ of a source galaxy at comoving distance $\chi_G$ and direction $\hat{\theta}$ is then related in the Born approximation to the matter density $\delta$ by the lensing kernel $W_L(z',z)$
\begin{equation}
\kappa(\hat{\theta})=\int_0^{\chi_G}\delta(\chi_L,\hat{\theta})W_L(\chi_L,\chi_G)d\chi_L.
\end{equation}
The 3D matter bispectrum is defined from $\kappa$ as
\begin{equation}
\langle \tilde{\kappa}(\bm{\ell_1})\tilde{\kappa}(\bm{\ell_2})\tilde{\kappa}(\bm{\ell_3})\rangle=(2\pi)^2\delta^{D}(\bm{\ell_1}+\bm{\ell_2}+\bm{\ell_3})B_{\delta}(\ell_1,\ell_2,\ell_3),\label{GIIeq:corr}
\end{equation}
where the ensemble average is denoted by $\langle\cdots\rangle$ and $\delta^{D}(\bm{\ell})$ is the Dirac delta function. $\delta^{D}(\bm{\ell}_1+\bm{\ell}_2+\bm{\ell}_3)$ enforces the condition that the three vectors form a triangle in Fourier space. Under the Limber approximation, we express the 2D angular cross-correlation bispectrum as
\begin{equation}
B_{ijk}^{\alpha\beta\gamma}(\ell;z^P_1,z^P_2,z^P_3)=\int_0^{\chi}\frac{W^{\alpha\beta\gamma}(\chi';\chi_1,\chi_2,\chi_3)}{\chi'^4}B_{\alpha\beta\gamma}(k;\chi')d\chi',\label{GIIeq:bs}
\end{equation}
where when $\alpha=\beta=\gamma=G$, for example, $B_{GGG}(k;\chi')$ is the 3D matter bispectrum shown in Eq. \ref{GIIeq:corr}. However, more generally $\alpha,\beta,\gamma\in G,I,g$, and the intrinsic alignment (I) and galaxy (g) bispectra are calculated as described below. The redshift is simply related through the Hubble parameter, $H(z)$, to $\chi$, and we can write the weighting function in terms of redshift as 
\begin{eqnarray}
W_{ijk}^{\alpha\beta\gamma}(z(\chi);z^P_1(\chi_1),z^P_2(\chi_2),z^P_3(\chi_3))&\equiv&W_{i}^{\alpha}(z,z^P_1)W_{j}^{\beta}(z,z^P_2)W_{k}^{\gamma}(z,z^P_3),\label{GIIeq:w}\\
W_{i}^G(z,z^P)&=&\int_0^{\infty}W_L(z',z)n^P_i(z^P)dz',\\
W_{i}^I(z,z^P)&=&W_{i}^g(z,z^P)=n^P_i(z^P).
\end{eqnarray}

$W_{i}^G(z,z^P)$ is simply the weighted lensing kernel for redshift bin $i$, while $W_{i}^I(z,z^P)=W_{i}^g(z,z^P)$ is the normalized galaxy photo-z distribution. The required lensing, intrinsic alignment, and galaxy bispectra are calculated following \cite{troxel} and \cite{troxel2}. A deterministic galaxy bias is assumed for the galaxy bispectra, and the galaxy intrinsic alignment signal is calculated using the fiducial parameterized model of \cite{SB10} (SB10), which is based on the halo model prescription. The SB10 model reduces by design to the linear alignment model \citep{hirata04} at large scale, but aims for a more motivated modelling of intrinsic alignment at small scales.

The flat-sky bispectrum is related to the all-sky bispectrum through the Wigner-3j symbol, where
\begin{equation}
B^{\alpha\beta\gamma}_{ijk,\ell_1\ell_2\ell_3}\approx\begin{pmatrix}\ell_1&\ell_2&\ell_3\\0&0&0\end{pmatrix}\sqrt{\frac{(2\ell_1+1)(2\ell_2+1)(2\ell_3+1)}{4\pi}}B^{\alpha\beta\gamma}_{ijk}(\ell_1,\ell_2,\ell_3).
\end{equation}
Following the fitting formula of \cite{sc} with coefficients $F^{\textrm{eff}}_2(k_1,k_2)$ described in Section 2.4.3 of \cite{tj}, we compute the 3D bispectrum due to nonlinear gravitational clustering
\begin{equation}
B_{\delta}(k_1,k_2,k_3;\chi)=2F^{\textrm{eff}}_2(k_1,k_2)P_{\delta}(k_1;\chi)P_{\delta}(k_2;\chi)+2\textrm{ perm.}
\end{equation}
We make a direct expansion of this method to approximate the 3D intrinsic alignment bispectra, using the intrinsic alignment power spectra instead of the nonlinear matter power spectrum, where
\begin{eqnarray}
B_{\delta\delta\tilde{\gamma}^I}(k_1,k_2,k_3;\chi)&=&2F^{\textrm{eff}}_2(k_1,k_2)P_{\delta\tilde{\gamma}^I}(k_1;\chi)P_{\delta}(k_2;\chi)+2F^{\textrm{eff}}_2(k_2,k_3)P_{\delta}(k_2;\chi)P_{\delta\tilde{\gamma}^I}(k_3;\chi)\\\nonumber
&&+2F^{\textrm{eff}}_2(k_3,k_1)P_{\delta\tilde{\gamma}^I}(k_3;\chi)P_{\delta\tilde{\gamma}^I}(k_1;\chi)\\
B_{\delta\tilde{\gamma}^I\tilde{\gamma}^I}(k_1,k_2,k_3;\chi)&=&2F^{\textrm{eff}}_2(k_1,k_2)P_{\tilde{\gamma}^I}(k_1;\chi)P_{\delta\tilde{\gamma}^I}(k_2;\chi)+2F^{\textrm{eff}}_2(k_2,k_3)P_{\delta\tilde{\gamma}^I}(k_2;\chi)P_{\delta\tilde{\gamma}^I}(k_3;\chi)\\\nonumber
&&+2F^{\textrm{eff}}_2(k_3,k_1)P_{\tilde{\gamma}^I}(k_3;\chi)P_{\delta\tilde{\gamma}^I}(k_1;\chi)\\
B_{\tilde{\gamma}^I}(k_1,k_2,k_3;\chi)&=&2F^{\textrm{eff}}_2(k_1,k_2)P_{\tilde{\gamma}^I}(k_1;\chi)P_{\tilde{\gamma}^I}(k_2;\chi)+2\textrm{ perm.}
\end{eqnarray}
This treatment provides reasonable results for the intrinsic alignment bispectra.

\section{The GII Self-Calibration Technique}\label{GIIsc}

A lensing survey will capture the information necessary for several sets of correlations between the $\delta_g$ and $\kappa^s$ which can be constructed for galaxy triplets. Like the GGI self-calibration technique, the GII self-calibration requires only three of these observed correlations. The first is the angular cross-correlation bispectrum between galaxy ellipticities ($\kappa^s$) in the \emph{i}-th, \emph{j}-th and \emph{k}-th redshift bins
\begin{equation}
B^{(1)}_{ijk}(\ell_1,\ell_2,\ell_3)=B^{GGG}_{ijk}(\ell_1,\ell_2,\ell_3)+B^{IGG}_{ijk}(\ell_1,\ell_2,\ell_3)+(\textrm{2 perm.})+B^{IIG}_{ijk}(\ell_1,\ell_2,\ell_3)+(\textrm{2 perm.})+B^{III}_{ijk}(\ell_1,\ell_2,\ell_3)\label{GIIeq:onea}.
\end{equation}
Unless catastrophic photo-z errors overwhelm the data, we can safely neglect the III cross-correlation, which is negligible for thick photo-z bins, by selecting galaxy triplets where $i<j<k$. Though the impact of the GII correlation under these conditions was neglected in \cite{troxel}, it can still impact the bispectrum for adjacent bins and should be taken into account if the goal is to make a precise measurement of the total intrinsic alignment contamination. For bin combinations which are not adjacent, this GII term can again be safely neglected and the GGI self-calibration is sufficient. Under this requirement, we still find that $B^{IIG}_{ijk}\gg B^{IGI}_{ijk},B^{GII}_{ijk}$ due to the lensing geometry. We then have for $i<j<k$,
\begin{equation}
B^{(1)}_{ijk}(\ell_1,\ell_2,\ell_3)\approx B^{GGG}_{ijk}(\ell_1,\ell_2,\ell_3)+B^{IGG}_{ijk}(\ell_1,\ell_2,\ell_3)+B^{IIG}_{ijk}(\ell_1,\ell_2,\ell_3).\label{GIIeq:one}
\end{equation}
The contribution from the GII bispectrum $B^{IIG}_{ijk}(\ell_1,\ell_2,\ell_3)$ ($i<j<k$) is typically very small and in some cases effectively zero. The GII self-calibration technique, which seeks to calculate and remove the GII cross-correlation which survives in adjacent photo-z bins, then acts as a correction to the GGI self-calibration, which is the only intrinsic alignment contamination that survives for non-adjacent photo-z bin combinations. We will denote the GII bispectrum $B^{IIG}_{ijk}$ in order to preserve the association of each quantity \emph{G} or \emph{I} to its redshift bin.

The second observable is the angular cross-correlation bispectrum between convergence ($\kappa^s$) in the \emph{i}-th redshift bin and the galaxy density ($\delta^{\Sigma}$) in the \emph{j}-th and \emph{k}-th redshift bins. The self-calibration requires the case where $i=j=k$
\begin{equation}
B^{(2)}_{iii}(\ell_1,\ell_2,\ell_3)=B^{GGg}_{iii}(\ell_1,\ell_2,\ell_3)+B^{GIg}_{iii}(\ell_1,\ell_2,\ell_3)+B^{IIg}_{iii}(\ell_1,\ell_2,\ell_3).\label{GIIeq:two}
\end{equation}
This correlation contributes further information about the intrinsic alignment of galaxies.

The final observable is measured in the angular cross-correlation bispectrum between galaxy density ($\delta^{\Sigma}$) in the \emph{i}-th, \emph{j}-th and \emph{k}-th redshift bins for $i=j=k$, giving
\begin{equation}
B^{(3)}_{iii}(\ell_1,\ell_2,\ell_3)=B^{ggg}_{iii}(\ell_1,\ell_2,\ell_3).\label{GIIeq:three}
\end{equation}
We also require for the GII self-calibration results from both the GI cross-correlation and intrinsic alignment auto-correlation self-calibration techniques \citep{22,23}. We have neglected thus far the contribution of magnification bias to these measurements. This was discussed and justified for the 3-point measurements in Sec. 4.3 of \cite{troxel} and was shown to be negligible. There is also a non-Gaussian contribution to the observed bispectra, which is also neglected as discussed in Sec. 4.4 of \cite{troxel}.

Our GII self-calibration technique calculates and removes the GII contamination in Eq. \ref{GIIeq:one} through the measurements from Eqs. \ref{GIIeq:two} \& \ref{GIIeq:three}, which are both available in the same lensing survey. The fractional contamination of the correlated intrinsic alignment to the lensing signal is expressed as
\begin{equation}
f^I_{ijk}(\ell_1,\ell_2,\ell_3)\equiv \frac{B^{IIG}_{ijk}(\ell_1,\ell_2,\ell_3)}{B^{GGG}_{ijk}(\ell_1,\ell_2,\ell_3)}.
\end{equation}
For the self-calibration to work, the contamination $f^I_{ijk}(\ell_1,\ell_2,\ell_3)$ must be sufficiently large as to contribute a detectable $B^{IIg}_{iii}(\ell_1,\ell_2,\ell_3)$ at the corresponding $\ell$ bins in $B^{(2)}_{iii}(\ell_1,\ell_2,\ell_3)$. We denote this threshold $f^{thresh}_{ijk}$. When $f^I_{ijk}\ge f^{thresh}_{ijk}$, the GII self-calibration can be applied to reduce the GII contamination. The residual error after the GII self-calibration will be expressed as a residual fractional error on the lensing measurement. We differentiate $\Delta f_{ijk}$ as statistical error and $\delta f_{ijk}$ as systematic error, as in the discussion of the GGI self-calibration. The performance of the GII self-calibration will then be quantified by the parameters $f^{thresh}_{ijk}$, $\Delta f_{ijk}$ and $\delta f_{ijk}$, which are discussed and calculated in Sec. \ref{GIIerror}. For parallelism, we preserve the notation from \cite{troxel}, though it is essential to note that the quantities are often not the same in each technique.

\subsection{Relating $B^{IIG}_{ijk}$ and $B^{IIg}_{iii}$}\label{GIIscaling}

In order to self-calibrate the GII bispectrum, we must first determine the relationship between $B^{IIG}_{ijk}$ and $B^{IIg}_{iii}$. We follow the same general approach as for the GGI bispectrum, relating the two by use of a deterministic galaxy bias $b_{g,k}$ \citep{26} such that the smoothed galaxy density is a function of matter density expressed as
\begin{equation}
\delta_g(\bm{x};\chi)=b_{g,1}(\chi)\delta_m(\bm{x};\chi)+\frac{b_{g,2}(\chi)}{2}\delta_m^2(\bm{x};\chi)+O(\delta^3).\label{GIIeq:density} 
\end{equation}
The linear galaxy bias ($b_{g,1}$) was used by \cite{22} for the 2-point correlations. In addition, $b_{g,2}$ represents the first order non-linear contribution. $b_{g,2}$ is typically expected to be negative and $\le b_{g,1}$ \citep{27}, and it is insufficient to model the bias as simply scale dependent as in the 2-point case \citep{28}. Following \cite{troxel}, we use this expression for the galaxy density to relate $B^{IIG}_{\delta\delta\gamma^I}$ to $B^{IIg}_{gg\gamma^I}$. We neglect the portion of the bispectrum due to primordial non-Gaussianity and the trispectrum term, which contains further information about the non-Gaussianity. This was justified and discussed further in Sec. 4.4 of \cite{troxel}, and results in the relationship
\begin{eqnarray}
B_{g\gamma^I\gamma^I}(k_1,k_2,k_3;\chi)&=&b_{g,1}(\chi)B_{\delta\gamma^I\gamma^I}(k_1,k_2,k_3;\chi)\nonumber\\
&&+\frac{b_{g,1}(\chi)b_{g,2}(\chi)}{2}\left[P_{\gamma^I\gamma^I}(k_1;\chi)P_{\delta\gamma^I}(k_2;\chi)+P_{\delta\gamma^I}(k_2;\chi)P_{\delta\gamma^I}(k_3;\chi)+P_{\gamma^I\gamma^I}(k_1;\chi)P_{\delta\gamma^I}(k_3;\chi)\right].\label{GIIeq:bg}
\end{eqnarray}

If the galaxy bias changes slowly over the \emph{i}-th redshift bin with median comoving distance $\chi_i$, we can write to a good approximation $b^i_k=b_{g,k}(\chi_i)$. We can further approximate $B(k_1,k_2,k_3;\chi)\approx B(k_1,k_2,k_3;\chi_i)$ and $P(k;\chi)\approx P(k;\chi_i)$ in the limit where the comoving distance distribution of galaxies in the \emph{i}-th redshift bin is narrow, and substituting Eq. \ref{GIIeq:bg} into Limber's approximation for $B^{IIg}_{iii}$ and comparing to $B^{IIG}_{iii}$, we have the approximations
\begin{eqnarray}
B^{IIg}_{iii}(\ell_1,\ell_2,\ell_3)&\approx & \frac{\Pi_{iii}}{\chi_i^4}\Big( b^{i}_{1}B_{\delta\gamma^I\gamma^I}(k_1,k_2,k_3;\chi_i)\nonumber\\
&&+\frac{b^{i}_{1}b^{i}_{2}}{2}\left[P_{\gamma^I\gamma^I}(k_1;\chi_i)P_{\delta\gamma^I}(k_2;\chi_i)+P_{\delta\gamma^I}(k_2;\chi_i)P_{\delta\gamma^I}(k_3;\chi_i)+P_{\gamma^I\gamma^I}(k_1;\chi_i)P_{\delta\gamma^I}(k_3;\chi_i)\right]\Big),\label{GIIeq:scale2b}
\end{eqnarray}
and
\begin{equation}
B^{IIG}_{ijk}(\ell_1,\ell_2,\ell_3)\approx B_{\delta\gamma^I\gamma^I}(k_1,k_2,k_3;\chi_i)\frac{W_{ijk}}{\chi_i^4}.\label{GIIeq:scale1b}
\end{equation}
where $W_{ijk}=\int_0^{\infty}f_i(\chi)f_j(\chi)W_k(\chi)d\chi$ and $\Pi_{iii}=\int_0^{\infty}f^3_i(\chi)d\chi$. From Eqs. \ref{GIIeq:scale1b} \& \ref{GIIeq:scale2b}, we can now write
\begin{eqnarray}
B^{IIG}_{ijk}(\ell_1,\ell_2,\ell_3)&\approx & \frac{W_{ijk}}{b^{i}_{1}\Pi_{iii}}B^{IIg}_{iii}(\ell_1,\ell_2,\ell_3)-\frac{b^{i}_{2}W_{ijk}}{\chi_i^4}\nonumber\\
&&\times\left[P_{\delta\gamma^I}(k_1;\chi_i)P_{\delta\delta}(k_2;\chi_i)+P_{\delta\delta}(k_2;\chi_i)P_{\delta\gamma^I}(k_3;\chi_i)+P_{\delta\gamma^I}(k_1;\chi_i)P_{\delta\gamma^I}(k_3;\chi_i)\right].\label{GIIeq:scale3}
\end{eqnarray}

In order to express the 3D power spectra in Eq. \ref{GIIeq:scale3} as 2D spectra, we will use the approximations $C^{Ig}_{ii}(\ell)\approx P_{\delta\gamma^I}(k;\chi_i)\frac{b^{i}_{1}\Pi_{ii}}{\chi^2_i}$ and $C^{II}_{ii}(\ell)\approx P_{\gamma^I\gamma^I}(k;\chi_i)\frac{\Pi_{ii}}{\chi^2_i}$, where $\Pi_{ii}=\int_0^{\infty}f^2_i(\chi)d\chi$. Equation \ref{GIIeq:scale3} is then
\begin{eqnarray}
B^{IIG}_{ijk}(\ell_1,\ell_2,\ell_3)&\approx & \frac{W_{ijk}}{b^{i}_{1}\Pi_{iii}}B^{IIg}_{iii}(\ell_1,\ell_2,\ell_3)-\frac{b^{i}_{2}}{b^{i}_{1}}\frac{W_{ijk}}{\Pi_{ii}^2}\nonumber\\
&&\times\left[C^{II}_{ii}(\ell_1)C^{Ig}_{ii}(\ell_2)+b^{i}_{1}C^{Ig}_{ii}(\ell_2)C^{Ig}_{ii}(\ell_3)+C^{II}_{ii}(\ell_1)C^{Ig}_{ii}(\ell_3)\right].\label{GIIeq:scale4}
\end{eqnarray}

This expression now relates $B^{IIG}_{ijk}$ to $B^{IIg}_{iii}$, allowing us to identify the GII bispectrum in the observable $B^{(2)}$. Like the similar expression for the GGI self-calibration, it is necessary to have information about the gI and II power spectra obtained from the 2-point self-calibration techniques \citep{22,23}.
\subsection{$B^{IIg}_{iii}$ estimator}\label{GIIcigg}
We can measure $B^{IIg}_{iii}$ through the information contained within the observable $B^{(2)}_{iii}=B^{IIg}_{iii}+B^{GIg}_{iii}+B^{GGg}_{iii}$. To measure it directly, we must first remove the contamination of the geometry dependent $B^{GIg}_{iii}+B^{Ggg}_{iii}$, which is not useful for the GII self-calibration. Lensing geometry simply requires eliminating those triplets of galaxies where the redshift of the galaxy used to measure the ellipticity is lower than those used to measure galaxy number density in the case of a spectroscopic galaxy sample. In this way, those triplets remaining have no contamination from lensing and measure only $B^{IIg}_{iii}$. However, for a photo-z galaxy sample, the typically large photo-z error prevents us from directly employing this method. Photo-z error causes a true redshift distribution of width $\ge 2\sigma_P=0.1(1+z)$ even for a photo-z bin with width $\Delta z\rightarrow 0$. In practice, photo-z bin widths are typically $\ge 0.2$. It is possible with such photo-z errors for galaxy triplets in the \emph{i}-th redshift bin to provide a measureable lensing contribution even when $i<j,k$, except for the special cases where the redshift or both the photo-z error and bin size are limited to sufficiently low values. A general photo-z galaxy sample then requires a more careful approach when separating $B^{IIg}_{iii}$ from $B^{GIg}_{iii}+B^{Ggg}_{iii}$.

In order to construct such an estimator, we first consider the orientation dependence of the two components. We define a redshift for each galaxy in the triplet: $z_G$, $z_{G'}$, $z_I$, or $z_{I'}$ for the galaxies used in the lensing/intrinsic alignment measurement and $z_g$ for the galaxy used in the number density measurement. The gII correlation is independent of the relative position of the three galaxies. For example, the correlations with $z_I < z_{I'} < z_{g}$, $z_g < z_I < z_{I'}$ or $z_{I'} < z_g < z_I$ are statistically identical when the sides of the triangle are fixed. However, the GGg and GIg correlations do depend on the relative position of the three galaxies. Due to the lensing geometry dependence, the correlation with $z_G , z_{G'/I} < z_g$ is statistically smaller than other orientations.

\begin{figure}
\center
\includegraphics[angle=270,scale=0.50]{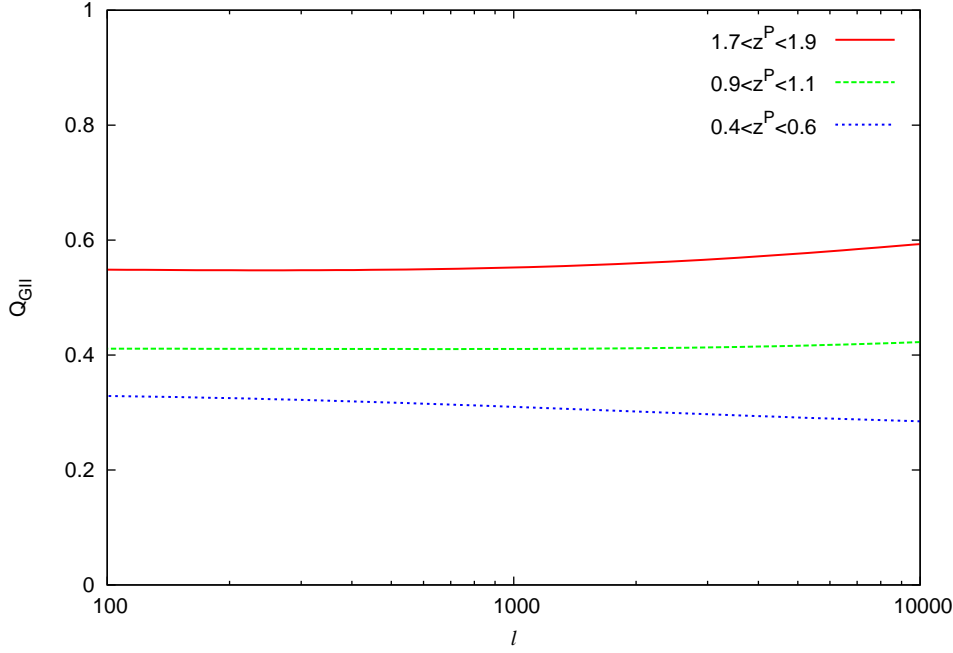}%
\caption{\label{GIIfig:q}The behavior of $Q_{GII}\equiv (B^{GGg}_{iii}|_S+B^{GIg}_{iii}|_S)/(B^{GGg}_{iii}+B^{GIg}_{iii})$ for equilateral triangles ($\ell=\ell_1=\ell_2=\ell_3$) over three redshift bins spanning the survey range. The expected values of $Q_{GII}$ are similar to those found for the GGI self-calibration, though with more $\ell$-dependence. This is primarily due to the greater complexity found in $Q_{GII}$, as the sum of two components instead of one. The suppression is dependent on the redshift bin chosen, increasing with redshift due to increased photo-z error at higher redshift. Generally, $Q\approx 0.4$, and the significant deviation from unity ensures that the estimator $\hat{B}^{IIg}_{iii}$ is valid for lensing surveys of interest.}
\end{figure}

This dependence provides two observables from $B^{(2)}_{iii}$. The first is $B^{(2)}_{iii}$, where all triplets are weighted equally. The second is $B^{(2)}_{iii}|_S$, which counts only those triplets with $z_G,z_{G'}< z_{g}$. This weighting is denoted by the subscript 'S'. This ordering is necessarily different from that employed in the GGI self-calibration. From our previous discussion, we then have $B^{IIg}_{iii}=B^{IIg}_{iii}|_S$ and $B^{GGg}_{iii}+B^{GIg}_{iii}> [B^{GGg}_{iii} +B^{GIg}_{iii}]|_S=B^{GGg}_{iii}|_S +B^{GIg}_{iii}|_S$. We now define the suppression ratio
\begin{equation}
Q_{GII}(\ell_1,\ell_2,\ell_3)\equiv\frac{B^{GGg}_{iii}|_S(\ell_1,\ell_2,\ell_3)+B^{GIg}_{iii}|_S(\ell_1,\ell_2,\ell_3)}{B^{GGg}_{iii}(\ell_1,\ell_2,\ell_3)+B^{GIg}_{iii}(\ell_1,\ell_2,\ell_3)},\label{GIIeq:qfact}
\end{equation}
where we have explicitly included the $\ell$-dependence which had been previously neglected. This ratio describes the suppression of the signal due to the weighting of triplets described for subscript 'S'. By definition $0< Q_{GII}< 1$, with $Q_{GII}=0$ if the photo-z is perfectly accurate and $Q_{GII}=1$ if the photo-z has no correlation to the true redshift. $Q_{GII}$ is calculated using the galaxy redshift distribution, which is discussed in Sec. \ref{GIIqfact}.

The definition of $Q_{GII}(\ell_1,\ell_2,\ell_3)$ possesses an explicit dependency on the intrinsic alignment, which in principle would make the estimator dependent on intrinsic alignment model. However, as we discuss in Sec. \ref{GIIqfact}, we find that for an LSST-like survey, $B^{GGg}_{iii}$ is large enough compared to $B^{GIg}_{iii}$ that we can approximate $Q_{GII}(\ell_1,\ell_2,\ell_3)\approx B^{GGg}_{iii}|_S/B^{GGg}_{iii}$ to very high accuracy. We can thus treat $Q_{GII}(\ell_1,\ell_2,\ell_3)$ as independent of intrinsic alignment model for sufficiently deep surveys.

We now define the estimator for $B^{IIg}_{iii}$, which we denote $\hat{B}^{IIg}_{iii}$, in terms of $Q_{GII}(\ell_1,\ell_2,\ell_3)$ and the two observables
\begin{eqnarray}
B^{(2)}_{iii}(\ell_1,\ell_2,\ell_3)&=&B^{IIg}_{iii}(\ell_1,\ell_2,\ell_3)+B^{GIg}_{iii}(\ell_1,\ell_2,\ell_3)+B^{GGg}_{iii}(\ell_1,\ell_2,\ell_3),\nonumber\\
B^{(2)}_{iii}|_S(\ell_1,\ell_2,\ell_3)&=&B^{IIg}_{iii}(\ell_1,\ell_2,\ell_3)+B^{GIg}_{iii}|_S(\ell_1,\ell_2,\ell_3)+B^{GGg}_{iii}|_S(\ell_1,\ell_2,\ell_3)\label{GIIeq:b2s}.
\end{eqnarray}
This estimator is
\begin{equation}
\hat{B}^{IIg}_{iii}(\ell_1,\ell_2,\ell_3)=\frac{B^{(2)}_{iii}|_S(\ell_1,\ell_2,\ell_3)-Q_{GII}(\ell_1,\ell_2,\ell_3)B^{(2)}_{iii}(\ell_1,\ell_2,\ell_3)}{1-Q_{GII}(\ell_1,\ell_2,\ell_3)}\label{GIIeq:cig}.
\end{equation}

As expected, when $Q_{GII}=0$ this gives $\hat{B}^{IIg}_{iii}=B^{(2)}_{iii}|_S$ as for a spectroscopic galaxy sample with no photo-z error. However, $Q_{GII}$ must not approach unity, where $\hat{B}^{IIg}_{iii}$ is singular. For the LSST-like survey described in Sec. \ref{GIIback}, we calculate $Q_{GII}$ for various redshift bins following the procedure described in Sec. \ref{GIIqfact}. This result is given in Fig. \ref{GIIfig:q} for equilateral triangles, where we find $Q_{GII}\approx 0.4$ and in general that $Q_{GII}$ should deviate significantly from unity. The estimator $\hat{B}^{IIg}_{iii}$ is thus expected to be applicable in any typical lensing survey.

\subsection{Evaluating $Q_{GII}(\ell_1,\ell_2,\ell_3)$}\label{GIIqfact}

The ratio $Q_{GII}$ in Eq. \ref{GIIeq:qfact} must first be evaluated in order to employ the estimator $\hat{B}^{IIg}_{iii}$. This mirrors the derivation in Sec. 3.3 of \cite{troxel}, which we will summarize here as it applies to $Q_{GII}$. We begin from the real space angular correlation functions $w^{GG'g}\left(\theta_1,\theta_2,\theta_3;z^P_G,z^P_{G'},z^P_g\right)$ and $w^{GIg}\left(\theta_1,\theta_2,\theta_3;z^P_G,z^P_{I},z^P_g\right)$. Taking the average correlation, we express this in terms of the ensemble average and calculate the associated bispectra. For the correlations with galaxy triplets weighted by the subscript 'S', we define the statistical weighting functions
\begin{eqnarray}
\eta(z_L,z_I,z_{g})&=&\frac{3\int_i dz^P_G\int_i dz^P_I\int_i dz^P_{g}\int_0^{\infty}dz_G W_L(z_L,z_G)N_i^P S(z^P_G,z^P_I,z^P_{g})}{\int_i dz^P_G\int_i dz^P_I\int_i dz^P_{g}\int_0^{\infty}dz_G W_L(z_L,z_G)N_i^P}\\\label{GIIeq:eta}
\eta'(z_L,z_{g'})&=&\frac{3\int_i dz^P_G\int_i dz^P_g\int_i dz^P_{g'}\int_0^{\infty}dz_G\int_0^{\infty}dz_{G'} W_L(z_L,z_G)W_L(z_L,z_{G'}){N'}_i^P S(z^P_G,z^P_{G'},z^P_{g})}{\int_i dz^P_G\int_i dz^P_g\int_i dz^P_{g'}\int_0^{\infty}dz_G\int_0^{\infty}dz_{G'} W_L(z_L,z_G)W_L(z_L,z_{G'}){N'}_i^P}
\end{eqnarray}
where
\begin{eqnarray}
N_i^P&\equiv& p(z_G|z_G^P)p(z_I|z_I^P)p(z_{g}|z_{g}^P)n_i^P(z^P_G)n_i^P(z^P_I)n_i^P(z^P_{g})\\
{N'}_i^P&\equiv& p(z_G|z_G^P)p(z_{G'}|z_{G'}^P)p(z_{g}|z_{g}^P)n_i^P(z^P_G)n_i^P(z^P_{G'})n_i^P(z^P_{g}),
\end{eqnarray}
and $S(z^P_G,z^P_I,z^P_{g})=1$ ($S(z^P_G,z^P_{G'},z^P_{g})=1$) if $z^P_G z^P_I< z^P_{g}$ ($z^P_G, z^P_{G'}< z^P_{g}$) and is zero otherwise. Since $S(\cdots)$ allow only $1/3$ of the integral to survive, $\eta$, $\eta'$ are normalised by a factor 3 in order to remove the suppression due to the selection function and measure only that due to the lensing geometry. 

We can now express directly the bispectra necessary to compute $Q_{GII}$
\begin{eqnarray}
B^{GIg}_{iii}(\ell_1,\ell_2,\ell_3)&=&\int_0^{\infty}B^{GIg}(k_1,k_2,k_3;\chi)\frac{W_i(\chi)f^2_i(\chi)}{\chi^4}d\chi\\
B^{GGg}_{iii}(\ell_1,\ell_2,\ell_3)&=&\int_0^{\infty}B^{GGg}(k_1,k_2,k_3;\chi)\frac{W^2_i(\chi)f_i(\chi)}{\chi^4}d\chi \label{GIIeq:q4}
\end{eqnarray}
and
\begin{eqnarray}
B^{GIg}_{iii}|_S(\ell_1,\ell_2,\ell_3)&=&\int_0^{\infty}B^{GIg}(k_1,k_2,k_3;\chi)\frac{W_i(\chi)f^2_i(\chi)}{\chi^4(z)}\eta(\chi,\chi(z_I)=\chi,\chi(z_{g})=\chi)d\chi\\
B^{GGg}_{iii}|_S(\ell_1,\ell_2,\ell_3)&=&\int_0^{\infty}B^{GGg}(k_1,k_2,k_3;\chi)\frac{W^2_i(\chi)f_i(\chi)}{\chi^4(z)}\eta'(\chi,\chi(z_g)=\chi)=\chi)d\chi. \label{GIIeq:q5}
\end{eqnarray}

The ratio $Q_{GII}$ is now defined through Eqs. \ref{GIIeq:q4} \& \ref{GIIeq:q5}. For the deep survey we are considering, we find that we can approximate to very high accuracy $Q_{GII}\approx\bar{\eta'}_i$, where $\bar{\eta'}_i$ is the mean value of $\eta'$ across the \emph{i}-th redshift bin. This is because for a deep survey, $B^{GGg}_{iii}$ is much larger than $B^{GIg}_{iii}$, making the contribution from $\eta$ negligible. In the limit where photo-z error dominates, $\sigma_P\gg\Delta z$, and so $\eta,\eta'\rightarrow 1$. In this limit, the estimator $\hat{B}^{IIg}_{iii}$ becomes singular and $B^{IIg}_{iii}$ can no longer be differentiated from $B^{GGg}_{iii}+B^{GIg}_{iii}$. In the opposite limit, where $\sigma_p\ll\Delta z$, $\eta,\eta'\rightarrow 0$, where our estimator mirrors the extraction method for $B^{IIg}_{iii}$ in spectroscopic galaxy samples. 
\section{Performance of the GII Self-Calibration}\label{GIIerror}
We will quantify the performance of the GII self-calibration technique using the survey parameters described in Sec. \ref{GIIback}. This includes both a statistical measurement error introduced through the estimator $\hat{B}^{IIg}_{iii}$, which propagates into the final measurement of $B^{IIG}_{ijk}$ and thus of $B^{GGG}_{ijk}$, and a systematic error due to inaccuracy in the scaling relation between $B^{IIG}_{ijk}$ and $B^{IIg}_{iii}$. We will also summarize other possible sources of error which might impact the performance of the self-calibration, though a more detailed account can be found in \cite{troxel}.

\subsection{The estimator $\hat{B}^{IIg}_{iii}$}\label{GIIciggerror}
The estimator $\hat{B}^{IIg}_{iii}$ introduces a statistical error into the measurement of $B^{IIg}_{iii}$ and thus $B^{GGG}_{ijk}$. In order to quantify the accuracy of the estimator $\hat{B}^{IIg}_{iii}(\ell_1,\ell_2,\ell_3)$, we consider the contribution of measurement errors such as shot and shape noise in $\hat{B}^{(2)}_{iii}(\ell_1,\ell_2,\ell_3)$ which propagate into our measurement of $B^{IIg}_{iii}(\ell_1,\ell_2,\ell_3)$ through the estimator. For a given redshift bin we calculate the rms error, working in a pixel space with $N_P$ sufficiently fine and uniform pixels of photo-z $z^P_{\alpha}$ and angular position $\theta_{\alpha}$. The measured overdensity is given by $\delta_{\alpha}+\delta^N_{\alpha}$, and the measured `shear' is given by $\kappa_{\alpha}+\kappa^I_{\alpha}+\kappa^N_{\alpha}$, where `N' represents the measurement noise. From Eq. \ref{GIIeq:b2s}, we construct the pixel space angular bispectra
\begin{eqnarray}
B^{(2)}(\ell_1,\ell_2,\ell_3)&=&N_P^{-3}\sum_{\alpha\beta\gamma}[\delta_{\alpha}+\delta^N_{\alpha}][\kappa_{\beta}+\kappa^I_{\beta}+\kappa^N_{\beta}][\kappa_{\gamma}+\kappa^I_{\gamma}+\kappa^N_{\gamma}]\exp[i(\bm{\ell_1}\cdot\bm{\theta_{\alpha}}+\bm{\ell_2}\cdot\bm{\theta_{\beta}}+\bm{\ell_3}\cdot\bm{\theta_{\gamma}})],\nonumber\\
B^{(2)}|_S(\ell_1,\ell_2,\ell_3)&=&N_P^{-3}\sum_{\alpha\beta\gamma}[\delta_{\alpha}+\delta^N_{\alpha}][\kappa_{\beta}+\kappa^I_{\beta}+\kappa^N_{\beta}][\kappa_{\gamma}+\kappa^I_{\gamma}+\kappa^N_{\gamma}]\exp[i(\bm{\ell_1}\cdot\bm{\theta_{\alpha}}+\bm{\ell_2}\cdot\bm{\theta_{\beta}}+\bm{\ell_3}\cdot\bm{\theta_{\gamma}})]S_{\alpha\beta\gamma}.\label{GIIeq:ester1}
\end{eqnarray}
$S_{\alpha\beta\gamma}=1$ when $z^P_{\alpha}>z^P_{\beta},z^P_{\gamma}$ and is zero otherwise. Thus in the limit $N_P\gg 1$, $\sum_{\alpha\beta\gamma}S_{\alpha\beta\gamma}=N_P^3/3$ and the average $\bar{S}_{\alpha\beta\gamma}=1/3$.

From our definition of the estimator in Eq. \ref{GIIeq:cig}, we can construct the difference
\begin{eqnarray}
\hat{B}^{IIg}_{iii}-B^{IIg}_{iii}&=&\frac{1}{(1-Q_{GII})}N_P^{-3}\sum_{\alpha\beta\gamma}\exp[i(\bm{\ell_1}\cdot\bm{\theta_{\alpha}}+\bm{\ell_2}\cdot\bm{\theta_{\beta}}+\bm{\ell_3}\cdot\bm{\theta_{\gamma}})][(3S_{\alpha\beta\gamma}-Q_{GII})\nonumber\\
&&\times(\delta_{\alpha}+\delta^N_{\alpha})(\kappa_{\beta}+\kappa^I_{\beta}+\kappa^N_{\beta})(\kappa_{\gamma}+\kappa^I_{\gamma}+\kappa^N_{\gamma})-(1-Q_{GII})\delta_{\alpha}\kappa^I_{\beta}\kappa^I_{\gamma}]\nonumber\\
&=&\frac{1}{(1-Q_{GII})}N_P^{-3}\sum_{\alpha\beta\gamma}\exp[i(\bm{\ell_1}\cdot\bm{\theta_{\alpha}}+\bm{\ell_2}\cdot\bm{\theta_{\beta}}+\bm{\ell_3}\cdot\bm{\theta_{\gamma}})](3S_{\alpha\beta\gamma}-Q_3)\nonumber\\
&&\times[(\delta_{\alpha}+\delta_{\alpha}^N)((\kappa_{\beta}+\kappa^I_{\beta}+\kappa^N_{\beta})(\kappa_{\gamma}+\kappa^N_{\gamma})+(\kappa_{\beta}+\kappa^N_{\beta})\kappa_{\gamma}^I)].\label{GIIeq:products}
\end{eqnarray}
We have used here that $\bar{S}_{\alpha\beta\gamma}=1/3$ and that the gII correlation doesn't depend on the relative position of the galaxy triplets. We can now write the rms error in a similar way to Eq. 41 of \cite{troxel} and simplify the resulting 256 6-point correlations which arise from expanding Eq. \ref{GIIeq:products}. We apply Wick's theorem, which allows us to express each 6-point correlation as 15 products of three 2-point correlations. This results in 3840 products, most of which are zero. For example, any correlation between signal and noise or dissimilar noise terms vanish. Further, due to the angular dependence of the correlations ($\langle A_{a}B_{b}\rangle=w_{AB}(\theta_{a}-\theta_{b})$), only those correlations with $\langle A_{a}B_{b}\rangle$ where \emph{a} $\in \alpha,\beta,\gamma$ and \emph{b} $\in \lambda,\mu,\nu$ are non-vanishing. This leaves 124 surviving products in the rms error expression
\begin{eqnarray}
\left(\Delta B^{IIg}_{iii}\right)^2&=&\frac{1}{\left(1-Q_{GII}\right)^2}N_P^{-6}\sum_{\alpha\beta\gamma}\sum_{\lambda\mu\nu}\exp\left[i(\bm{\ell_1}\cdot\bm{\theta_{\alpha}}+\bm{\ell_2}\cdot\bm{\theta_{\beta}}+\bm{\ell_3}\cdot\bm{\theta_{\gamma}})\right]\exp\left[i(\bm{\ell_1}\cdot\bm{\theta_{\lambda}}+\bm{\ell_2}\cdot\bm{\theta_{\mu}}+\bm{\ell_3}\cdot\bm{\theta_{\nu}})\right]\left(3S_{\alpha\beta\gamma}-Q_3\right)\nonumber\\
&&\times\left(3S_{\lambda\mu\nu}-Q_3\right)\Big\{\big(\langle\kappa_{\beta}\kappa_{\mu}\rangle+\langle\kappa_{\beta}^N\kappa_{\mu}^N\rangle\big)\Big[\big(\langle\delta_{\alpha}\delta_{\lambda}\rangle+\langle\delta_{\alpha}^N\delta_{\lambda}^N\rangle\big)\big(\langle(\kappa_{\gamma}+I_{\gamma})(\kappa_{\nu}+I_{\nu})\rangle+\langle\kappa_{\gamma}^N\kappa_{\nu}^N\rangle\big)\nonumber\\
&&+\langle\delta_{\alpha}(\kappa_{\nu}+I_{\nu})\rangle\langle(\kappa_{\gamma}+I_{\gamma})\delta_{\lambda}\rangle\Big] 
+\big(\langle\kappa_{\beta}\kappa_{\nu}\rangle+\langle\kappa_{\beta}^N\kappa_{\nu}^N\rangle\big)\Big[\big(\langle\delta_{\alpha}\delta_{\lambda}\rangle+\langle\delta_{\alpha}^N\delta_{\lambda}^N\rangle\big)\big(\langle(\kappa_{\gamma}+I_{\gamma})(\kappa_{\mu}+I_{\mu})\rangle+\langle\kappa_{\gamma}^N\kappa_{\mu}^N\rangle\big)\nonumber\\
&&+\langle\delta_{\alpha}(\kappa_{\mu}+I_{\mu})\rangle\langle(\kappa_{\gamma}+I_{\gamma})\delta_{\lambda}\rangle\Big]
+\big(\langle\kappa_{\gamma}\kappa_{\mu}\rangle+\langle\kappa_{\gamma}^N\kappa_{\mu}^N\rangle\big)\Big[\big(\langle\delta_{\alpha}\delta_{\lambda}\rangle+\langle\delta_{\alpha}^N\delta_{\lambda}^N\rangle\big)\big(\langle(\kappa_{\beta}+I_{\beta})I_{\nu}\rangle+\langle I_{\beta}\kappa_{\nu}\rangle\big)\nonumber\\
&&+\langle\delta_{\alpha}(\kappa_{\nu}+I_{\nu})\rangle\langle(\kappa_{\beta}+I_{\beta})\delta_{\lambda}\rangle\Big]
+\big(\langle\kappa_{\gamma}\kappa_{\nu}\rangle+\langle\kappa_{\gamma}^N\kappa_{\nu}^N\rangle\big)\Big[\big(\langle\delta_{\alpha}\delta_{\lambda}\rangle+\langle\delta_{\alpha}^N\delta_{\lambda}^N\rangle\big)\big(\langle(\kappa_{\beta}+I_{\beta})I_{\mu}\rangle+\langle I_{\beta}\kappa_{\mu}\rangle\big)\\
&&+\langle\delta_{\alpha}(\kappa_{\mu}+I_{\mu})\rangle\langle(\kappa_{\beta}+I_{\beta})\delta_{\lambda}\rangle\Big]
+\big(\langle\delta_{\alpha}\delta_{\lambda}\rangle+\langle\delta_{\alpha}^N\delta_{\lambda}^N\rangle\big)\Big[\langle\kappa_{\beta}I_{\mu}\rangle\langle I_{\gamma}\kappa_{\nu}\rangle+\langle\kappa_{\beta}I_{\nu}\rangle\langle I_{\gamma}\kappa_{\mu}\rangle+\langle\kappa_{\gamma}I_{\mu}\rangle\langle I_{\beta}\kappa_{\nu}\rangle\nonumber\\
&&+\langle\kappa_{\gamma}I_{\nu}\rangle\langle I_{\beta}\kappa_{\mu}\rangle\Big]
+\langle\kappa_{\beta}\delta_{\lambda}\rangle\Big[\langle\delta_{\alpha}\kappa_{\mu}\rangle\langle(\kappa_{\gamma}+I_{\gamma})I_{\nu}\rangle+\langle\delta_{\alpha}(\kappa_{\mu}+I_{\mu})\rangle\langle I_{\gamma}\kappa_{\nu}\rangle+\langle\delta_{\alpha}\kappa_{\nu}\rangle\langle(\kappa_{\gamma}+I_{\gamma})I_{\mu}\rangle\nonumber\\
&&+\langle\delta_{\alpha}(\kappa_{\nu}+I_{\nu})\rangle\langle I_{\gamma}\kappa_{\mu}\rangle\Big]
+\langle\kappa_{\gamma}\delta_{\lambda}\rangle\Big[\langle\delta_{\alpha}\kappa_{\mu}\rangle\langle(\kappa_{\beta}+I_{\beta})I_{\nu}\rangle+\langle\delta_{\alpha}\kappa_{\nu}\rangle\langle(\kappa_{\beta}+I_{\beta})I_{\mu}\rangle+\langle\delta_{\alpha}(\kappa_{\nu}+I_{\nu})\rangle\langle I_{\beta}\kappa_{\mu}\rangle\nonumber\\
&&+\langle\delta_{\alpha}(\kappa_{\mu}+I_{\mu})\rangle\langle I_{\beta}\kappa_{\nu}\rangle\Big]+\langle I_{\beta}\delta_{\lambda}\rangle\Big[\langle\delta_{\alpha}\kappa_{\mu}\rangle\langle\kappa_{\gamma}I_{\nu}\rangle+\langle\delta_{\alpha}\kappa_{\nu}\rangle\langle\kappa_{\gamma}I_{\mu}\rangle\Big]+\langle I_{\gamma}\delta_{\lambda}\rangle\Big[\langle\delta_{\alpha}\kappa_{\mu}\rangle\langle\kappa_{\beta}I_{\nu}\rangle+\langle\delta_{\alpha}\kappa_{\nu}\rangle\langle\kappa_{\beta}I_{\mu}\rangle\Big]\Big\}.\nonumber\label{GIIeq:ester2}
\end{eqnarray}

Noises only correlate at zero lag ($\langle\delta^N_{\alpha}\delta^N_{\lambda}\rangle\propto \delta_{\alpha\lambda}$, $\langle\kappa^N_{\gamma}\kappa^N_{\nu}\rangle\propto \delta_{\gamma\nu}$), and the correlations not involving $\kappa$ depend only on separation, not on relative orientation of the galaxy pairs along the line-of-sight. However, correlations like $\langle\kappa\delta\rangle$ and $\langle\kappa\kappa^I\rangle$ are dependent on the relative orientation and must be treated with care when evaluating Eq. \ref{GIIeq:ester2}. In order to quantify this orientation dependence, we apply $Q_{Ig}\equiv Q_2$ \citep{22,troxel} and $Q_{IG}$ such that
\begin{eqnarray}
\langle\delta_{\alpha}\kappa_{\nu}\rangle\rightarrow\frac{1}{2} \left(\frac{S_{\alpha\nu}}{(1-Q_{Ig})}+\frac{S_{\nu\alpha}}{Q_{Ig}}\right)\langle\delta_{\alpha}\kappa_{\nu}\rangle\nonumber\\
\langle\kappa^I_{\beta}\kappa_{\mu}\rangle\rightarrow\frac{1}{2} \left(\frac{S_{\beta\mu}}{(1-Q_{IG})}+\frac{S_{\mu\beta}}{Q_{IG}}\right)\langle\kappa^I_{\beta}\kappa_{\mu}\rangle.\label{GIIeq:sub}
\end{eqnarray}
\begin{figure}
\center
\includegraphics[angle=270,scale=0.5]{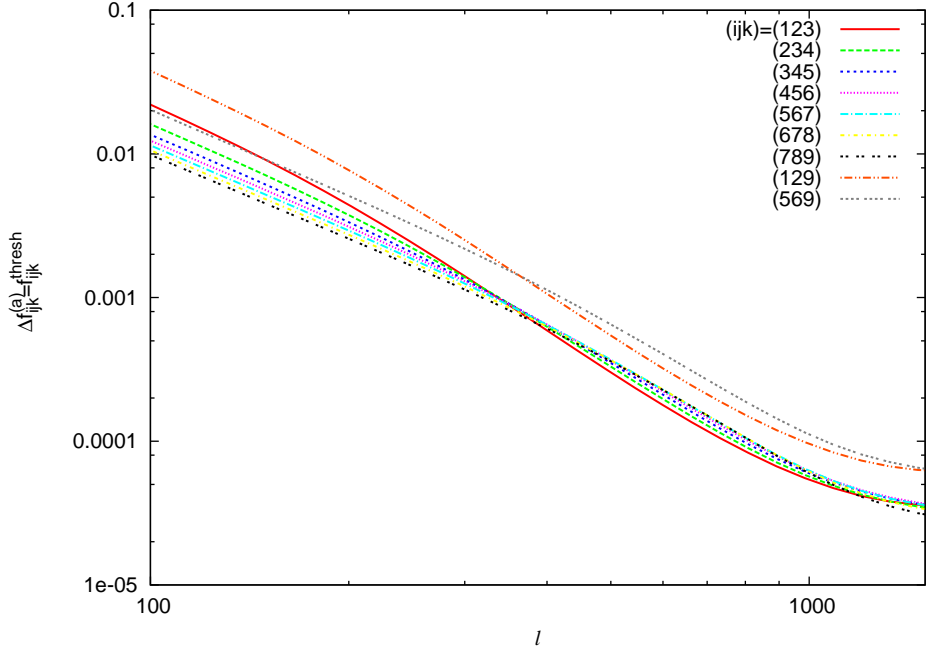}%
\caption{\label{GIIfig:merror}The residual statistical uncertainty $\Delta f^{(a)}_{ijk}$ for equilateral triangles ($\ell=\ell_1=\ell_2=\ell_3$) in the $B^{IIG}_{ijk}$ measurement and threshold of intrinsic alignment contamination $f^{thresh}_{ijk}$ at which the GII self-calibration technique can calculate and remove the intrinsic alignment contamination at S/N=1 are plotted for a variety of redshift bin combinations. At large $\ell$, we see the effects of shot noise beginning to take over. Generally, $\Delta f^{(a)}_{ijk}$ is less than the minimum survey error expected for such a survey, and is thus negligible. We expect this result to hold for non-equilateral triangles as well, but the use of the GII self-calibration is limited by our understanding of non-Gaussian effects for very elongated triangle shapes, as discussed in \protect\cite{troxel}, and we leave discussion of its applicability for these very elongated triangle shapes to a future work.}
\end{figure}

The suppression ratio $Q_{IG}\approx Q_{Ig}$ is defined in an identical way to $Q_{Ig}$, but for the $C^{IG}$ power spectrum. We can now evaluate Eq. \ref{GIIeq:ester2} analytically, converting to its Fourier representation
\begin{eqnarray}
\left(\Delta B^{IIg}_{iii}\right)^2&=&2\Big\{\left(C^{\text{GG}} +C^{\text{II}}\right)C^{\text{Gg}}C^{\text{Gg}}K_1+C^{\text{GG}} \big[\left(C^{\text{GG}} +2C^{\text{II}}\right)\left( C^{\text{gg}}+ C^{\text{ggN}} a_1\right)+2C^{\text{Ig}}C^{\text{Ig}}\nonumber\\
&&+ C^{\text{Gg}}C^{\text{Ig}}K_2+C^{\text{IG}} \left(C^{\text{gg}} K_3+C^{\text{ggN}} K_4\right)+C^{\text{GGN}} \left(C^{\text{gg}} K_5+C^{\text{ggN}} K_6\right)\big] \nonumber\\
&&+C^{\text{IG}} \big[C^{\text{IG}} \left(C^{\text{gg}}K_7+C^{\text{ggN}} K_8\right)+C^{\text{Gg}}C^{\text{Gg}} K_9+ C^{\text{Gg}}C^{\text{Ig}}K_{10}\\
&&+C^{\text{GGN}} \left(C^{\text{gg}} K_{11}+C^{\text{ggN}} K_{12}\right)\big]\nonumber\\
&&+C^{\text{GGN}} \big[C^{\text{GGN}} \left(C^{\text{gg}} c_1+ C^{\text{ggN}} e_1\right)+ C^{\text{Ig}}C^{\text{Ig}} K_5+C^{\text{II}} \left(C^{\text{gg}} K_5+C^{\text{ggN}} K_6\right)\nonumber\\
&&+ C^{\text{Gg}}C^{\text{Gg}}K_{13}+ C^{\text{Gg}}C^{\text{Ig}} K_{14}\big]\Big\}
\nonumber\label{GIIeq:ester3}
\end{eqnarray}
The details of this calculation and the coefficients $K_1-K14$ and $a_1,c_1,e_1$ are included in the appendix.

The final rms error $\Delta B^{IIg}_{iii}$ evaluated for a given triangle with bin width $\Delta \ell$ is then given by
\begin{eqnarray}
\left(\Delta B^{IIg}_{iii}\right)^2&=&\frac{4\pi^2}{\ell_1\ell_2\ell_3\Delta\ell_1\Delta\ell_2\Delta\ell_3f_{sky}}\Big\{\left(C^{\text{GG}} +C^{\text{II}}\right)C^{\text{Gg}}C^{\text{Gg}}K_1\nonumber\\
&&+C^{\text{GG}} \big[\left(C^{\text{GG}} +2C^{\text{II}}\right)\left( C^{\text{gg}}+ C^{\text{ggN}} a_1\right)+2C^{\text{Ig}}C^{\text{Ig}}\nonumber\\
&&+ C^{\text{Gg}}C^{\text{Ig}}K_2+C^{\text{IG}} \left(C^{\text{gg}} K_3+C^{\text{ggN}} K_4\right)+C^{\text{GGN}} \left(C^{\text{gg}} K_5+C^{\text{ggN}} K_6\right)\big] \nonumber\\
&&+C^{\text{IG}} \big[C^{\text{IG}} \left(C^{\text{gg}}K_7+C^{\text{ggN}} K_8\right)+C^{\text{Gg}}C^{\text{Gg}} K_9+ C^{\text{Gg}}C^{\text{Ig}}K_{10}\\
&&+C^{\text{GGN}} \left(C^{\text{gg}} K_{11}+C^{\text{ggN}} K_{12}\right)\big]\nonumber\\
&&+C^{\text{GGN}} \big[C^{\text{GGN}} \left(C^{\text{gg}} c_1+ C^{\text{ggN}} e_1\right)+ C^{\text{Ig}}C^{\text{Ig}} K_5+C^{\text{II}} \left(C^{\text{gg}} K_5+C^{\text{ggN}} K_6\right)\nonumber\\
&&+ C^{\text{Gg}}C^{\text{Gg}}K_{13}+ C^{\text{Gg}}C^{\text{Ig}} K_{14}\big]\Big\}\nonumber\label{GIIeq:ester4}
\end{eqnarray}
$C^{gg,N}_{ii}=1/\bar{n}_i$ and $C^{GG,N}_{ii}=\gamma^2_{rms}/\bar{n}_i$, where $\bar{n}_i$ is the average number density of galaxies in the \emph{i}-th redshift bin. The statistical error $\Delta B^{IIg}_{iii}$ differs in several key ways from the statistical error $\Delta B^{IIg}_{iii}$ in the GGI self-calibration, both in the complexity of its calculation due to the additional $\kappa$ term and the resulting components, which include terms like $C^{IG}$. It is more sensitive to the intrinsic alignment contamination in the limit than in the Igg and 2-point cases, but is still safe from strong dependence in the limit where $C^{GG}_{ii}$ is dominant.

The errors $\Delta B^{IIg}_{iii}$ and $\Delta C^{Ig}_{ii}$ \citep{22} propagate into the measurement of $B^{IIG}_{iii}$ through Eq. \ref{GIIeq:scale4}. To find the fractional error $\Delta f^{(a)}_{ijk}$ this induces in the lensing bispectrum, we simply scale $\Delta B^{IIG}_{ijk}$ by the factor $f^I_{ijk}$ such that $\Delta f^{(a)}_{ijk}=f^I_{ijk}\Delta B^{IIG}_{ijk}$. This error is equal to $f^{thresh}_{ijk}$, the minimum intrinsic alignment $f^I_{ijk}$ which can be detected through the self-calibration with S/N=1 or $\Delta B^{IIg}_{iii}=B^{IIg}_{iii}$. Thus both the residual statistical error in the measurement of $C^{GGG}_{ijk}$ and the lower limit at which the intrinsic alignment can be calculated and removed with the self-calibration is represented by $f^{thresh}_{ijk}=\Delta f^{(a)}_{ijk}$. Thus the GII self-calibration technique can turn a systematic contamination $f^I_{ijk}$ of the lensing signal into a statistical error $\Delta f^{(a)}_{ijk}<f^I_{ijk}$, which is insensitive to the original intrinsic alignment contamination.

We show in Fig. \ref{GIIfig:merror} $\Delta f^{(a)}_{ijk}$ for the survey parameters described in Sec. \ref{GIIback}. Compared to the minimum survey error as found in \cite{troxel}, $\Delta f^{(a)}_{ijk}$ is negligible for most scales, only becoming comparable for some photo-z bin combinations at very large scale.

\begin{figure}
\center
\includegraphics[angle=270,scale=0.45]{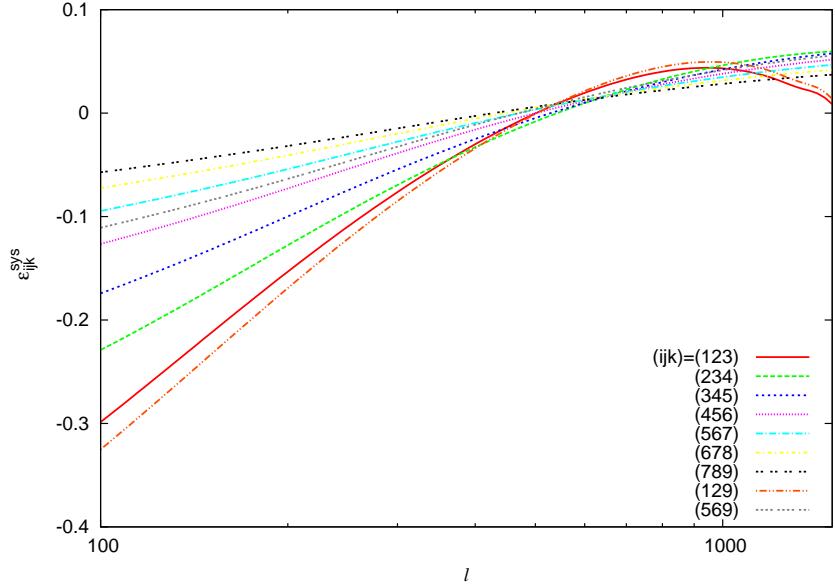}%
\caption{\label{GIIfig:serror}The inaccuracy of the relationship between $B^{IIG}_{ijk}$ and the observable $B^{IIg}_{iii}$ is quantified in Eq. \ref{GIIeq:sys1} by $\epsilon^{sys}_{ijk}$. The dominant systematic error in the measurement of $B^{GGG}_{ijk}$ through the GII self-calibration technique is due to this inaccuracy. For equilateral triangles ($\ell=\ell_1=\ell_2=\ell_3$), $\epsilon^{sys}_{ijk}$ is plotted for several sets of adjacent redshift bins, where the stronger dependence of the lensing kernel on redshift causes a significantly higher inaccuracy. We neglect totally non-adjacent photo-z bin combinations, as the GII signal is truly negligible in such cases. Equation \ref{GIIeq:scale4} is accurate to within $\sim 30\%$ for all scales and photo-z bin combinations. However, for most scales and photo-z bin choices, it is accurate to within $10\%$. This is comparable to the accuracy found in the GGI self-calibration, and we expect the GII self-calibration to be capable of reducing the GII intrinsic alignment contamination by a factor of 10 or so in general. These results are insensitive to the original intrinsic alignment contamination, such that for any $f^{thresh}_{ijk}<f^I_{ijk}<1$, the GGI self-calibration will reduce the GGI contamination down to survey limits or by a factor of 10 or so, whichever is less, for all but a few photo-z bin choices at large scale.}
\end{figure}
\subsection{The accuracy of the $B^{IIG}_{ijk}$-$B^{IIg}_{iii}$ relation}\label{GIIscalingerror}
In addition to the statistical error introduced through the estimator $\hat{B}^{IIg}_{iii}$, there is a systematic error which is introduced by Eq. \ref{GIIeq:scale4}, which relates the intrinsic alignment contamination $B^{IIG}_{ijk}$ in the lensing bispectrum to other survey observables. The accuracy of Eq. \ref{GIIeq:scale4} is quantified by
\begin{eqnarray}
\epsilon^{sys}_{ijk}&\equiv & \Bigg(\frac{W_{ijk}}{(b^{i}_{1})^2\Pi_{iii}}\frac{B^{IIg}_{iii}(\ell_1,\ell_2,\ell_3)}{B^{IIG}_{ijk}(\ell_1,\ell_2,\ell_3)}-\frac{b^{i}_{2}}{(b^{i}_{1})^2}\frac{W_{ijk}}{\omega_{ii}\Pi_{ii}}\frac{1}{B^{IIG}_{ijk}(\ell_1,\ell_2,\ell_3)}\nonumber\\
&&\times\left[C^{Ig}_{ii}(\ell_1)C^{GG}_{ii}(\ell_2)+C^{GG}_{ii}(\ell_2)C^{Ig}_{ii}(\ell_3)+\frac{\omega_{ii}}{b^i_1\Pi_{ii}}C^{Ig}_{ii}(\ell_1)C^{Ig}_{ii}(\ell_3)\right]\Bigg)^{-1}-1.\label{GIIeq:sys1}
\end{eqnarray}
This induces a residual systematic error in the lensing measurement of
\begin{equation}
\delta f_{ijk}=\epsilon^{sys}_{ijk}f^I_{ijk}.\label{GIIeq:sys2}
\end{equation}

$\epsilon^{sys}_{ijk}$ is evaluated numerically and shown in Fig. \ref{GIIfig:serror} for equilateral triangles and a variety of neighboring photo-z bin choices. We neglect totally non-adjacent photo-z bin combinations, as the GII signal is truly negligible in such cases. The accuracy of Eq. \ref{GIIeq:scale4} reflects the differences between $B^{IIG}_{ijk}$ and $B^{IGG}_{ijk}$. The accuracy of Eq. \ref{GIIeq:scale4} depends primarily on mean redshift of the photo-z bins chosen, becoming more accurate at higher redshift where the intrinsic alignment is less strong. By comparison, the accuracy of the equivalent scaling relation for the GGI self-calibration shows strong tendencies toward the effects of the lensing kernel, since the lensing contribution to the signal is much stronger. Equation \ref{GIIeq:scale4} is accurate to within $\sim 30\%$ for all scales and photo-z bin combinations. However, for most scales and photo-z bin choices, it is accurate to within $10\%$. This is comparable to the accuracy found in the GGI self-calibration, and we expect the GII self-calibration to be capable of reducing the GII intrinsic alignment contamination by a factor of 10 or so in general. These results are insensitive to the original intrinsic alignment contamination, such that for any $f^{thresh}_{ijk}<f^I_{ijk}<1$, the GGI self-calibration will reduce the GGI contamination down to survey limits or by a factor of 10 or so, whichever is less, for all but a few photo-z bin choices at large scale. The actual impact of the systematic on the measurement of $B^{GGG}$ is significantly less, however, as it is scaled by $f^I_{ijk}$, which is typically expected to be on the order of 10-20\%. This would lead to an actual systematic error on the percent level.
\subsection{Other sources of uncertainty}\label{GIIother}

The GII self-calibration suffers from many additional potential sources of uncertainty, including uncertainties in galaxy bias modelling, limitations in the models used for the intrinsic alignment and bispectrum calculations, magnification bias effects, and the effects of non-Gaussianity in the bispectrum. \cite{troxel} provides detailed calculations which compare the order at which magnification and non-Gaussian effects, as well as uncertainties in the galaxy bias model, might impact the self-calibration for the GGI bispectrum. While not all of these calculations apply directly to the GII self-calibration, it is trivial to replicate them following \cite{troxel}. They also address a range of other possible factors which might influence the self-calibration, including for example cosmological uncertainties and catastrophic photo-z errors. In all cases, they find that the dominant impact on the performance of the self-calibration comes from the statistical and systematic errors described in Secs. \ref{GIIciggerror} and \ref{GIIscalingerror}. We find that this holds for the GII self-calibration as well, with the exception of uncertainty in the galaxy bias parameters, which becomes comparable to the statistical measurement error in $B^{IIG}$ at large scale.
\subsection{Summary of residual errors}\label{GIIbehavior}
The performance of the GII self-calibration can be summarised under three regimes, which are defined by the magnitude of the GII contamination as represented by $f^I_{ijk}$. The first is where the gII correlation is too small to detect in $B^{(2)}$, with $f^I_{ijk}\le f^{thresh}_{ijk}$. If the intrinsic alignment cannot be detected in $B^{(2)}$, the GII self-calibration is not applicable. This generally means that the GII contamination is also negligible when compared to $\epsilon^{min}_{ijk}$, the minimum statistical error in the lensing bispectrum, and there is no need to correct for it. This is likely true for all totally non-adjacent photo-z bin choices, where the GII signal is naturally negligible as discussed in \cite{troxel3}.

However, the GII contamination to the lensing bispectrum is likely not negligible if $f^I_{ijk}>f^{thresh}_{ijk}$ and it must be corrected for. The GII self-calibration is now able to detect and calculate the GII cross-correlation. In the second regime, where $\Delta f^{thresh}_{ijk}>\epsilon^{sys}_{ijk}f^I_{ijk}$, the statistical error $\Delta f^{(a)}_{ijk}$ induced by measurement error in the estimator $\hat{B}^{IIg}_{iii}$ is dominant. This error is generally negligible when compared to $\epsilon^{min}_{ijk}$, and so in this regime, the GII self-calibration should perform at the statistical limit of the lensing survey.

Where $\Delta f^{thresh}_{ijk}<\epsilon^{sys}_{ijk}f^I_{ijk}$, the systematic error $\delta f_{ijk}=\epsilon^{sys}_{ijk}f^I_{ijk}$ due to the relationship between $B^{IIG}_{ijk}$ and $B^{IIg}_{iii}$ in Eq. \ref{GIIeq:scale4} is dominant. In the case where $\epsilon^{sys}_{ijk}<\epsilon^{min}_{ijk}/f^I_{ijk}$, $\epsilon^{min}_{ijk}$ is still dominant. Otherwise the GII self-calibration can suppress the GII contamination by a factor of 10 or so for all but a few photo-z bin choices at the largest scales. In this case, other complementary techniques could be employed to further reduce the GII contamination down to the statistical limit for the lensing survey.

For example, one such case has been explored by \cite{23} for the 2-point correlations, but such studies of the 3-point intrinsic alignment are left to be done. \cite{36} combines the GI self-calibration with a photo-z self-calibration to better protect the GI self-calibration against catastrophic photo-z effects. Both methods are possible because the GI and GGI self-calibration uses primarily those correlations in one redshift bin to estimate the intrinsic alignment, while \cite{23,36} use those correlations between redshift bins. Others have also used information between redshift bins to calibrate the intrinsic alignment contamination in the 2- and 3-point correlations \citep{14b,19,20a,20b,21,9b}. Such techniques for the 3-point intrinsic alignment correlations should eventually complement the GII and GGI self-calibration techniques for improved reductions in the contamination by the intrinsic alignment in the cosmic shear signal, but further exploration is necessary in order to jointly apply these complementary techniques to realistic survey conditions.
\section{Conclusion}\label{GIIconc}
The 3-point intrinsic alignment correlations (GGI, GII, III) are expected to strongly contaminate the galaxy lensing bispectrum at up to the 20\% level. \cite{troxel3} showed that while the III correlation can be safely neglected by considering only the cross-correlation bispectrum between three different photo-z bins, the GII cross-correlation remains a contaminant for adjacent bins in addition to the GGI cross-correlation and thus must be considered in any self-calibration of adjacent photo-z bin combinations. \cite{troxel} first generalized the self-calibration technique to the bispectrum in order to calculate and remove the 3-point GGI contamination from the GGG bispectrum. In this work we extend the self-calibration to the 3-point GII cross-correlation in order to measure and remove its remaining contamination from adjacent bin triplets. 

In order to do this, we establish the estimator $\hat{B}^{IIg}_{iii}$ to extract the gII correlation from the galaxy ellipticity-ellipticity-density measurement for a photo-z galaxy sample. This estimator is expected to be generally applicable to weak lensing surveys and reduces to the simple extraction method for spectroscopic galaxy samples at low photo-z error. We then develop a scaling relation between the GII and gII bispectra using the linear and non-linear galaxy bias to relate the galaxy density and cosmic shear measurements. We can then calculate and remove the GII correlation from the GGG bispectrum. While this method is in principle applicable to all $\ell$ and triangle shapes, we maintain the modest restrictions of \cite{troxel} on very elongated triangles due to the effects of non-Gaussianity and at very non-linear scales due to limitations in the understanding of the galaxy bias model used. Combining the GII and GGI self-calibration techniques will then allow a complete removal of the 3-point intrinsic alignment contamination from the cosmic shear signal.

The performance of the GII self-calibration technique is also quantified for a typical weak-lensing survey. The residual statistical error due to measurement uncertainty in the estimator $\hat{B}^{IIg}_{iii}$ is shown to be generally negligible when compared to the minimum measurement error in the lensing bispectrum. We consider the systematic error introduced by the relationship between $B^{IIG}_{ijk}$ and $B^{IIg}_{iii}$, showing that $|\epsilon_{ijk}|<0.3$ for all photo-z bin choices and all scales, while except for the very largest scales, $|\epsilon_{ijk}|<0.1$. The intrinsic alignment contamination can then be reduce the GII contamination by a factor of 10 or more for all adjacent photo-z bin combinations at $\ell>300$. For larger scales, we find that the GII contamination can be reduced by a factor of 3-5 or more. This will potentially allow the GII self-calibration to reduce the GII correlation to the statistical limit of the lensing survey, as discussed in Sec. \ref{GIIbehavior}.

These results are not strongly sensitive to the original intrinsic alignment contamination, such that for any $f^{thresh}_{ijk}<f^I_{ijk}<1$, the GII self-calibration can reduce the GII contamination down to survey limits or by a factor of 10 or so, whichever is less, for all but the largest scales. This is comparable to the GGI self-calibration, where for any $f^{thresh}_{ij}<f^I_{ij}<1$, it can reduces the GGI contamination down to survey limits or by a factor of 10 or greater for most photo-z bin choices, whichever is less. We thus expect the GII self-calibration to perform well with the GGI self-calibration, and together they promise to be an efficient technique to isolate the total 3-point intrinsic alignment signal from the cosmic shear measurement.

\section*{Acknowledgments}
The authors would like to thank P. Zhang for valuable comments. MI acknowledges that this material is based upon work supported in part by National Science Foundation under grant AST-1109667 and NASA under grant NNX09AJ55G, and that part of the calculations for this work have been performed on the Cosmology Computer Cluster funded by the Hoblitzelle Foundation.
\appendix
\section{Calculation of coefficients in $\Delta B^{IIg}_{\MakeLowercase{iii}}$}

When evaluating the sum and converting Eq. \ref{GIIeq:ester2} to Fourier space, the products of the correlations each have a numerical coefficient due to the restrictions on redshift ordering. Many, however, are identical due to symmetries, and others are numerically equivalent when specific suppression ratios are chosen. The calculation of the unique coefficients $K_1-K16$ and $a_1,c_1,e_1$ in Eqs. \ref{GIIeq:ester3} \& \ref{GIIeq:ester4} are summarised here. The coefficients $K_1-K16$ are simply collections of other coefficients, as shown
\begin{eqnarray}
K_1 &\equiv & a_3+b_3\nonumber\\
K_2 &\equiv & a_2+b_2+2c_2\nonumber\\
K_3 &\equiv & d_2+e_2+f_2+g_2\nonumber\\
K_4 &\equiv & h_2+i_2+j_2+k_2\nonumber\\
K_5 &\equiv & a_1+b_1\nonumber\\
K_6 &\equiv & c_1+d_1\nonumber\\
K_7 &\equiv & g_3+h_3\\
K_8 &\equiv & i_3+j_3\nonumber\\
K_9 &\equiv & a_4+b_4+c_4+d_4\nonumber\\
K_{10} &\equiv & c_3+d_3+e_3+f_3\nonumber\\
K_{11} &\equiv & h_2+n_2+o_2+(j_2+q_2)/2\nonumber\\
K_{12} &\equiv & r_2+s_2+t_2+u_2\nonumber\\
K_{13} &\equiv & k_3+l_3\nonumber\\
K_{14} &\equiv & l_2+m_2+q_2+r_2\nonumber
\end{eqnarray}

From these, we group the coefficients by the number of orientation dependent correlations, for example $\langle\delta\kappa\rangle$, are involved in their calculation. The first coefficient is trivial, due to products with no noise correlations or orientation dependent correlations. We then calculate 
\begin{eqnarray}
\frac{N_P^{-6}}{(1-Q_3)^2}\sum_{\alpha\beta\gamma}\sum_{\lambda\mu\nu}(3S_{\alpha\beta\gamma}-Q_3)(3S_{\lambda\mu\nu}-Q_3)\approx 1.\label{GIIeq:appendix}
\end{eqnarray}
For the following terms, each is calculated as in the case of Eq. \ref{GIIeq:appendix}, but limited by some ordering restriction due to a noise or orientation dependent correlation. For those terms with no orientation dependent correlations, there exist five unique coefficients:
\begin{eqnarray}
a_1&\equiv &1+\frac{7}{20 \left(1-Q_{\text{GII}}\right){}^2}\\
b_1&\equiv &1-\frac{7}{40 \left(1-Q_{\text{GII}}\right){}^2}\\
c_1&\equiv &1+\frac{1}{8 \left(1-Q_{\text{GII}}\right){}^2}\\
d_1&\equiv &1+\frac{7}{8 \left(1-Q_{\text{GII}}\right){}^2}\\
e_1&\equiv &1+\frac{1}{2 \left(1-Q_{\text{GII}}\right){}^2}.
\end{eqnarray}
The coefficient $a_1$ is achieved through the restrictions $\delta_{\alpha\lambda}$, $\delta_{\beta\nu}$, and $\delta_{\gamma\nu}$; $b_1$ is achieved through $\delta_{\beta\mu}$ and $\delta_{\gamma\mu}$; $c_1$ through $\delta_{\alpha\lambda}\delta_{\beta\mu}$, $\delta_{\alpha\lambda}\delta_{\gamma\mu}$, $\delta_{\beta\mu}\delta_{\gamma\nu}$, and $\delta_{\beta\nu}\delta_{\gamma\mu}$; $d_1$ through $\delta_{\alpha\lambda}\delta_{\beta\nu}$ and $\delta_{\alpha\lambda}\delta_{\gamma\nu}$; and finally $e_1$ through $\delta_{\alpha\lambda}\delta_{\beta\mu}\delta_{\gamma\nu}$ and $\delta_{\alpha\lambda}\delta_{\beta\nu}\delta_{\gamma\mu}$.

For those with one orientation dependent correlations, there exist 24 unique coefficients:
\begin{eqnarray}
a_2&\equiv &1+\frac{29-18 Q_{\text{Ig}}}{40 \left(1-Q_{\text{GII}}\right){}^2}-\frac{11-6 Q_{\text{Ig}}}{8 \left(1-Q_{\text{GII}}\right)}\\
b_2&\equiv &1+\frac{17+6 Q_{\text{Ig}}}{40 \left(1-Q_{\text{GII}}\right){}^2}-\frac{11-6 Q_{\text{Ig}}}{8 \left(1-Q_{\text{GII}}\right)}\\
c_2&\equiv &1+\frac{49-18 Q_{\text{Ig}}}{80 \left(1-Q_{\text{GII}}\right){}^2}-\frac{11-6 Q_{\text{Ig}}}{8 \left(1-Q_{\text{GII}}\right)}\\
d_2&\equiv &1+\frac{61-42 Q_{\text{IG}}}{80 \left(1-Q_{\text{GII}}\right){}^2}-\frac{1}{1-Q_{\text{GII}}}\\
e_2&\equiv &1+\frac{7+6 Q_{\text{IG}}}{20 \left(1-Q_{\text{GII}}\right){}^2}-\frac{1}{1-Q_{\text{GII}}}\\
f_2&\equiv &1+\frac{19+42 Q_{\text{IG}}}{80 \left(1-Q_{\text{GII}}\right){}^2}-\frac{1}{1-Q_{\text{GII}}}\\
g_2&\equiv &1+\frac{13-6 Q_{\text{IG}}}{20 \left(1-Q_{\text{GII}}\right){}^2}-\frac{1}{1-Q_{\text{GII}}}\\
h_2&\equiv &1-\frac{3 \left(-7+5 Q_{\text{IG}}\right)}{20 \left(1-Q_{\text{GII}}\right){}^2}-\frac{1}{1-Q_{\text{GII}}}\\
i_2&\equiv &1+\frac{3 \left(7+4 Q_{\text{IG}}\right)}{40 \left(1-Q_{\text{GII}}\right){}^2}-\frac{1}{1-Q_{\text{GII}}}\\
j_2&\equiv &1+\frac{3 \left(2+5 Q_{\text{IG}}\right)}{20 \left(1-Q_{\text{GII}}\right){}^2}-\frac{1}{1-Q_{\text{GII}}}\\
k_2&\equiv &1-\frac{3 \left(-11+4 Q_{\text{IG}}\right)}{40 \left(1-Q_{\text{GII}}\right){}^2}-\frac{1}{1-Q_{\text{GII}}}\\
l_2&\equiv &1+\frac{3 \left(5+Q_{\text{Ig}}\right)}{40 \left(1-Q_{\text{GII}}\right){}^2}-\frac{11-6 Q_{\text{Ig}}}{8 \left(1-Q_{\text{GII}}\right)}\\
m_2&\equiv &1-\frac{3 \left(-8+5 Q_{\text{Ig}}\right)}{40 \left(1-Q_{\text{GII}}\right){}^2}-\frac{11-6 Q_{\text{Ig}}}{8 \left(1-Q_{\text{GII}}\right)}\\
n_2&\equiv &1+\frac{3 \left(3+5 Q_{\text{IG}}\right)}{40 \left(1-Q_{\text{GII}}\right){}^2}-\frac{1}{1-Q_{\text{GII}}}\\
o_2&\equiv &1-\frac{3 \left(-8+5 Q_{\text{IG}}\right)}{40 \left(1-Q_{\text{GII}}\right){}^2}-\frac{1}{1-Q_{\text{GII}}}\\
p_2&\equiv &1-\frac{9 \left(-2+Q_{\text{Ig}}\right)}{20 \left(1-Q_{\text{GII}}\right){}^2}-\frac{11-6 Q_{\text{Ig}}}{8 \left(1-Q_{\text{GII}}\right)}\\
q_2&\equiv &1+\frac{7+6 Q_{\text{IG}}}{20 \left(1-Q_{\text{GII}}\right){}^2}-\frac{1}{1-Q_{\text{GII}}}\\
r_2&\equiv &1+\frac{3 \left(1+Q_{\text{IG}}\right)}{8 \left(1-Q_{\text{GII}}\right){}^2}-\frac{1}{1-Q_{\text{GII}}}\\
s_2&\equiv &1-\frac{3 \left(-2+Q_{\text{IG}}\right)}{8 \left(1-Q_{\text{GII}}\right){}^2}-\frac{1}{1-Q_{\text{GII}}}
\end{eqnarray}
\begin{eqnarray}
t_2&\equiv &1+\frac{3 \left(1+3 Q_{\text{IG}}\right)}{8 \left(1-Q_{\text{GII}}\right){}^2}-\frac{1}{1-Q_{\text{GII}}}\\
u_2&\equiv &1+\frac{14-13 Q_{\text{IG}}}{8 \left(1-Q_{\text{GII}}\right){}^2}-\frac{7+2 Q_{\text{IG}}}{8 \left(1-Q_{\text{GII}}\right)}.
\end{eqnarray}
\begin{table}
\renewcommand{\arraystretch}{1.5}
\center
\begin{tabular}{ c c c c c c c c c c }
\hline
 Coeff. & Value &Coeff. & Value &Coeff. & Value &Coeff. & Value &Coeff. & Value  \\
\hline
$a_1$ & $\frac{71}{36}\approx 1.97$ & $e_2$ & $\frac{13}{18}\approx 0.72$ & $n_2$ & $\frac{23}{48}\approx 0.48$ & $a_3$ & $\frac{11}{6}\approx 1.83$ & $i_3$ & $\frac{119}{144}\approx 0.83$ \\
$b_1$ & $\frac{37}{72}\approx 0.54$ & $f_2$ & $\frac{13}{18}\approx 0.72$ & $o_2$ & $\frac{23}{48}\approx 0.48$ & $b_3$ & $\frac{79}{144}\approx 0.56$ & $j_3$ & $\frac{119}{144}\approx 0.83$\\
$c_1$ & $\frac{97}{72}\approx 1.35$ & $g_2$ & $\frac{13}{18}\approx 0.72$ & $p_2$ & $\frac{29}{24}\approx 1.21$ & $c_3$ & $\frac{7}{12}\approx 0.58$ & $k_3$ & $\frac{133}{288}\approx 0.46$  \\
$d_1$ & $\frac{247}{72}\approx 3.43$ & $h_2$ & $\frac{29}{24}\approx 1.21$ & $q_2$ & $\frac{13}{18}\approx 0.72$ & $d_3$ & $\frac{7}{12}\approx 0.58$ & $l_3$ & $\frac{119}{144}\approx 0.83$ \\
$e_1$ & $\frac{43}{18}\approx 2.39$ & $i_2$ & $\frac{29}{24}\approx 1.21$ & $r_2$ & $\frac{43}{48}\approx 0.90$ & $e_3$ & $\frac{7}{12}\approx 0.58$ &  $a_4$ & $\frac{37}{72}\approx 0.51$  \\
$a_2$ & $\frac{13}{18}\approx 0.72$ & $j_2$ & $\frac{29}{24}\approx 1.21$ & $s_2$ & $\frac{43}{48}\approx 0.90$ & $f_3$ & $\frac{7}{12}\approx 0.58$ &  $b_4$ & $\frac{37}{72}\approx 0.51$    \\
$b_2$ & $\frac{13}{18}\approx 0.72$ & $k_2$ & $\frac{29}{24}\approx 1.21$ & $t_2$ & $\frac{31}{16}\approx 1.94$ & $g_3$ & $\frac{7}{12}\approx 0.58$ & $c_4$ & $\frac{37}{72}\approx 0.51$ \\
$c_2$ & $\frac{13}{18}\approx 0.72$ & $l_2$ & $\frac{23}{48}\approx 0.48$ & $u_2$ & $\frac{31}{16}\approx 1.94$ & $h_3$ & $\frac{7}{12}\approx 0.58$ & $d_4$ & $\frac{37}{72}\approx 0.51$ \\
$d_2$ & $\frac{13}{18}\approx 0.72$ & $m_2$ & $\frac{23}{48}\approx 0.48$ & & & & \\\hline
\end{tabular}
\caption{Numerical values for coefficients in the calculation of Eqs. \ref{GIIeq:ester3} \& \ref{GIIeq:ester4} for typical suppression ratios $Q_{GII}\approx 2/5$, $Q_{IG}\approx Q_{Ig}\approx 1/2$.}
\label{GIItable}
\end{table}
Using the shorthand $\chi^{AB}_{\alpha\beta}\equiv\left(S_{\alpha\beta}/(1-Q_{AB})+S_{\beta\alpha}/Q_{AB}\right)/2$ for Eq. \ref{GIIeq:sub}, we can express the coefficient conditions as follows. The coefficient $a_2$ is achieved through the restriction $\chi^{Ig}_{\alpha\mu}$; $b_2$ is achieved through $\chi^{Ig}_{\alpha\nu}$; $c_2$ through $\chi^{Ig}_{\lambda\beta}$ and $\chi^{Ig}_{\lambda\gamma}$; $d_2$ through $\chi^{IG}_{\beta\mu}$ and $\chi^{IG}_{\gamma\mu}$; $e_2$ through $\chi^{IG}_{\beta\nu}$ and $\chi^{IG}_{\gamma\nu}$; $f_2$ through $\chi^{IG}_{\mu\beta}$ and $\chi^{IG}_{\mu\gamma}$; $g_2$ through $\chi^{IG}_{\nu\beta}$ and $\chi^{IG}_{\nu\gamma}$; $h_2$ through $\delta_{\alpha\lambda}\chi^{IG}_{\beta\mu}$, $\delta_{\alpha\lambda}\chi^{IG}_{\gamma\mu}$, $\delta_{\gamma\nu}\chi^{IG}_{\beta\mu}$, and $\delta_{\beta\nu}\chi^{IG}_{\gamma\mu}$; $i_2$ through $\delta_{\alpha\lambda}\chi^{IG}_{\beta\nu}$ and $\delta_{\alpha\lambda}\chi^{IG}_{\gamma\nu}$; $j_2$ through $\delta_{\alpha\lambda}\chi^{IG}_{\mu\beta}$ and $\delta_{\alpha\lambda}\chi^{IG}_{\mu\gamma}$, and $\delta_{\gamma\nu}\chi^{Ig}_{\mu\beta}$; $k_2$ through $\delta_{\alpha\lambda}\chi^{IG}_{\nu\beta}$ and $\delta_{\alpha\lambda}\chi^{IG}_{\nu\gamma}$; $l_2$ through $\delta_{\beta\mu}\chi^{Ig}_{\alpha\nu}$ and $\delta_{\gamma\mu}\chi^{Ig}_{\alpha\nu}$; $m_2$ through $\delta_{\beta\mu}\chi^{Ig}_{\lambda\gamma}$ and $\delta_{\gamma\mu}\chi^{Ig}_{\lambda\beta}$; $n_2$ through $\delta_{\beta\mu}\chi^{IG}_{\gamma\nu}$ and $\delta_{\gamma\mu}\chi^{IG}_{\beta\nu}$; $o_2$ through $\delta_{\beta\mu}\chi^{IG}_{\nu\gamma}$ and $\delta_{\gamma\mu}\chi^{IG}_{\nu\beta}$; $p_2$ through $\delta_{\beta\nu}\chi^{Ig}_{\alpha\mu}$ and $\delta_{\gamma\nu}\chi^{Ig}_{\alpha\mu}$; $q_2$ through $\delta_{\beta\nu}\chi^{Ig}_{\lambda\gamma}$, $\delta_{\gamma\nu}\chi^{Ig}_{\lambda\beta}$, and $\delta_{\beta\nu}\chi^{IG}_{\mu\gamma}$; $r_2$ through $\delta_{\alpha\lambda}\delta_{\beta\mu}\chi^{IG}_{\gamma\nu}$ and $\delta_{\alpha\lambda}\delta_{\gamma\mu}\chi^{IG}_{\beta\nu}$; $s_2$ through $\delta_{\alpha\lambda}\delta_{\beta\mu}\chi^{IG}_{\nu\gamma}$ and $\delta_{\alpha\lambda}\delta_{\gamma\mu}\chi^{IG}_{\nu\beta}$; $t_2$ through $\delta_{\alpha\lambda}\delta_{\beta\nu}\chi^{IG}_{\mu\gamma}$ and $\delta_{\alpha\lambda}\delta_{\gamma\nu}\chi^{IG}_{\mu\beta}$; and finally $u_2$ through $\delta_{\alpha\lambda}\delta_{\beta\nu}\chi^{IG}_{\gamma\mu}$ and $\delta_{\alpha\lambda}\delta_{\gamma\nu}\chi^{IG}_{\beta\mu}$.

For those with two orientation dependent correlations, there exist 12 unique coefficients:
\begin{eqnarray}
a_3&\equiv &1-\frac{3 \left(-24-41 Q_{\text{Ig}}+50 Q_{\text{Ig}}^2\right)}{80 \left(1-Q_{\text{GII}}\right){}^2}-\frac{37-13 Q_{\text{Ig}}-2 Q_{\text{Ig}}^2}{20 \left(1-Q_{\text{GII}}\right)}\\
b_3&\equiv &1+\frac{64-27 Q_{\text{Ig}}+28 Q_{\text{Ig}}^2}{80 \left(1-Q_{\text{GII}}\right){}^2}-\frac{72-25 Q_{\text{Ig}}-3 Q_{\text{Ig}}^2}{40 \left(1-Q_{\text{GII}}\right)}\\
c_3&\equiv &1-\frac{3 \left(-23-Q_{\text{Ig}}+3 Q_{\text{IG}}+8 Q_{\text{Ig}} Q_{\text{IG}}\right)}{80 \left(1-Q_{\text{GII}}\right){}^2}-\frac{68-16 Q_{\text{Ig}}-Q_{\text{IG}}+2 Q_{\text{Ig}} Q_{\text{IG}}}{40 \left(1-Q_{\text{GII}}\right)}\\
d_3&\equiv &1+\frac{3 \left(20-4 Q_{\text{Ig}}+3 Q_{\text{IG}}+2 Q_{\text{Ig}} Q_{\text{IG}}\right)}{80 \left(1-Q_{\text{GII}}\right){}^2}-\frac{68-16 Q_{\text{Ig}}-Q_{\text{IG}}+2 Q_{\text{Ig}} Q_{\text{IG}}}{40 \left(1-Q_{\text{GII}}\right)}\\
e_3&\equiv &1+\frac{56-13 Q_{\text{Ig}}+2 Q_{\text{IG}}+38 Q_{\text{Ig}} Q_{\text{IG}}}{80 \left(1-Q_{\text{GII}}\right){}^2}-\frac{68-16 Q_{\text{Ig}}-Q_{\text{IG}}+2 Q_{\text{Ig}} Q_{\text{IG}}}{40 \left(1-Q_{\text{GII}}\right)}\\
f_3&\equiv &1+\frac{35-4 Q_{\text{Ig}}-Q_{\text{IG}}-10 Q_{\text{Ig}} Q_{\text{IG}}}{40 \left(1-Q_{\text{GII}}\right){}^2}-\frac{68-16 Q_{\text{Ig}}-Q_{\text{IG}}+2 Q_{\text{Ig}} Q_{\text{IG}}}{40 \left(1-Q_{\text{GII}}\right)}\\
g_3&\equiv &1+\frac{3 \left(16+5 Q_{\text{IG}}+6 Q_{\text{IG}}^2\right)}{80 \left(1-Q_{\text{GII}}\right){}^2}-\frac{31-4 Q_{\text{IG}}+4 Q_{\text{IG}}^2}{20 \left(1-Q_{\text{GII}}\right)}\\
h_3&\equiv &1+\frac{3 \left(27-17 Q_{\text{IG}}+6 Q_{\text{IG}}^2\right)}{80 \left(1-Q_{\text{GII}}\right){}^2}-\frac{31-4 Q_{\text{IG}}+4 Q_{\text{IG}}^2}{20 \left(1-Q_{\text{GII}}\right)}\\
i_3&\equiv &1+\frac{25+13 Q_{\text{IG}}+8 Q_{\text{IG}}^2}{40 \left(1-Q_{\text{GII}}\right){}^2}-\frac{31-4 Q_{\text{IG}}+4 Q_{\text{IG}}^2}{20 \left(1-Q_{\text{GII}}\right)}\\
j_3&\equiv &1+\frac{46-29 Q_{\text{IG}}+8 Q_{\text{IG}}^2}{40 \left(1-Q_{\text{GII}}\right){}^2}-\frac{31-4 Q_{\text{IG}}+4 Q_{\text{IG}}^2}{20 \left(1-Q_{\text{GII}}\right)}\\
k_3&\equiv &1+\frac{34-17 Q_{\text{Ig}}+11 Q_{\text{Ig}}^2}{40 \left(1-Q_{\text{GII}}\right){}^2}-\frac{37-13 Q_{\text{Ig}}-2 Q_{\text{Ig}}^2}{20 \left(1-Q_{\text{GII}}\right)}\\
l_3&\equiv &1+\frac{37-5 Q_{\text{Ig}}-4 Q_{\text{Ig}}^2}{40 \left(1-Q_{\text{GII}}\right){}^2}-\frac{37-13 Q_{\text{Ig}}-2 Q_{\text{Ig}}^2}{20 \left(1-Q_{\text{GII}}\right)}.
\end{eqnarray}
The coefficient $a_3$ is achieved through the restrictions $\chi^{Ig}_{\alpha\mu}\chi^{Ig}_{\lambda\beta}$ and $\chi^{Ig}_{\alpha\mu}\chi^{Ig}_{\lambda\gamma}$; $b_3$ is achieved through $\chi^{Ig}_{\alpha\nu}\chi^{Ig}_{\lambda\beta}$ and $\chi^{Ig}_{\alpha\nu}\chi^{Ig}_{\lambda\gamma}$; $c_3$ through $\chi^{Ig}_{\lambda\gamma}\chi^{IG}_{\beta\mu}$ and $\chi^{Ig}_{\lambda\beta}\chi^{IG}_{\gamma\mu}$; $d_3$ through $\chi^{Ig}_{\lambda\gamma}\chi^{IG}_{\beta\nu}$ and $\chi^{Ig}_{\lambda\beta}\chi^{IG}_{\gamma\nu}$; 
$e_3$ through $\chi^{Ig}_{\alpha\nu}\chi^{IG}_{\mu\beta}$ and $\chi^{Ig}_{\alpha\nu}\chi^{IG}_{\mu\gamma}$;
$f_3$ through $\chi^{Ig}_{\alpha\mu}\chi^{IG}_{\nu\beta}$ and $\chi^{Ig}_{\alpha\mu}\chi^{IG}_{\nu\gamma}$;
$g_3$ through $\chi^{IG}_{\gamma\nu}\chi^{IG}_{\mu\beta}$ and $\chi^{IG}_{\beta\nu}\chi^{IG}_{\mu\gamma}$;
$h_3$ through $\chi^{IG}_{\gamma\mu}\chi^{IG}_{\nu\beta}$ and $\chi^{IG}_{\beta\mu}\chi^{IG}_{\nu\gamma}$;
$i_3$ through $\chi^{IG}_{\gamma\nu}\chi^{IG}_{\mu\beta}$ and $\chi^{IG}_{\beta\nu}\chi^{IG}_{\mu\gamma}$;
$j_3$ through $\chi^{IG}_{\nu\gamma}\chi^{IG}_{\beta\mu}$ and $\chi^{IG}_{\nu\beta}\chi^{IG}_{\gamma\mu}$;
$k_3$ through $\delta_{\beta\mu}\chi^{Ig}_{\alpha\nu}\chi^{Ig}_{\lambda\gamma}$ and $\delta_{\gamma\mu}\chi^{Ig}_{\alpha\nu}\chi^{Ig}_{\lambda\beta}$;
and finally $l_3$ through $\delta_{\beta\nu}\chi^{Ig}_{\alpha\mu}\chi^{Ig}_{\lambda\gamma}$ and $\delta_{\gamma\nu}\chi^{Ig}_{\alpha\mu}\chi^{Ig}_{\lambda\beta}$.

Finally, for those with three orientation dependent correlations, there exist 4 unique coefficients:
\begin{eqnarray}
a_4&\equiv &1+\frac{74-9 Q_{\text{Ig}}+23 Q_{\text{Ig}}^2-11 Q_{\text{Ig}} Q_{\text{IG}}-20 Q_{\text{Ig}}^2 Q_{\text{IG}}}{80 \left(1-Q_{\text{GII}}\right){}^2}-\frac{77-13 Q_{\text{Ig}}-2 Q_{\text{Ig}}^2}{40 \left(1-Q_{\text{GII}}\right)}\\
b_4&\equiv &1+\frac{74-20 Q_{\text{Ig}}+3 Q_{\text{Ig}}^2+11 Q_{\text{Ig}} Q_{\text{IG}}+20 Q_{\text{Ig}}^2 Q_{\text{IG}}}{80 \left(1-Q_{\text{GII}}\right){}^2}-\frac{77-13 Q_{\text{Ig}}-2 Q_{\text{Ig}}^2}{40 \left(1-Q_{\text{GII}}\right)}\\
c_4&\equiv &1+\frac{75-12 Q_{\text{Ig}}-8 Q_{\text{Ig}}^2+2 Q_{\text{IG}}+3 Q_{\text{Ig}} Q_{\text{IG}}+10 Q_{\text{Ig}}^2 Q_{\text{IG}}}{80 \left(1-Q_{\text{GII}}\right){}^2}-\frac{77-13 Q_{\text{Ig}}-2 Q_{\text{Ig}}^2}{40 \left(1-Q_{\text{GII}}\right)}\\
d_4&\equiv &1+\frac{77-9 Q_{\text{Ig}}+2 Q_{\text{Ig}}^2-2 Q_{\text{IG}}-3 Q_{\text{Ig}} Q_{\text{IG}}-10 Q_{\text{Ig}}^2 Q_{\text{IG}}}{80 \left(1-Q_{\text{GII}}\right){}^2}-\frac{77-13 Q_{\text{Ig}}-2 Q_{\text{Ig}}^2}{40 \left(1-Q_{\text{GII}}\right)}.
\end{eqnarray}
The coefficient $a_4$ is achieved through the restrictions $\chi^{Ig}_{\alpha\nu}\chi^{Ig}_{\lambda\beta}\chi^{IG}_{\gamma\mu}$ and $\chi^{Ig}_{\alpha\nu}\chi^{Ig}_{\lambda\gamma}\chi^{IG}_{\beta\mu}$; $b_4$ is achieved through $\chi^{Ig}_{\alpha\nu}\chi^{Ig}_{\lambda\beta}\chi^{IG}_{\mu\gamma}$ and $\chi^{Ig}_{\alpha\nu}\chi^{Ig}_{\lambda\gamma}\chi^{IG}_{\mu\beta}$; $c_4$ through $\chi^{Ig}_{\alpha\mu}\chi^{Ig}_{\lambda\beta}\chi^{IG}_{\gamma\nu}$ and $\chi^{Ig}_{\alpha\mu}\chi^{Ig}_{\lambda\gamma}\chi^{IG}_{\beta\nu}$; and finally $d_4$ through $\chi^{Ig}_{\alpha\mu}\chi^{Ig}_{\lambda\beta}\chi^{IG}_{\nu\gamma}$ and $\chi^{Ig}_{\alpha\mu}\chi^{Ig}_{\lambda\gamma}\chi^{IG}_{\nu\beta}$. 

For typical suppression ratios $Q_{GII}\approx 2/5$, $Q_{IG}=Q_{Ig}=1/2$, this produces the numerical values given in Table A1.

\label{lastpage}

\end{document}